\documentclass[english]{cccconf}
\usepackage[comma,numbers,square,sort&compress]{natbib}
\usepackage{amsmath}
\usepackage{subfigure}
\usepackage{algorithm}
\usepackage{algorithmic}
\newtheorem{myDef}{Definition}

\begin{document}
\begin{sloppypar}

\title{Distributed processing of continuous range queries over moving objects}
\author{Hui Zhu\aref{amss}, 
        Ziqiang Yu\aref{amss}}



\affiliation[amss]{Yantai University
        \email{zqyu@ytu.edu.cn, wumie698@163.com}}
\maketitle

\begin{abstract}
Monitoring range queries over moving objects is essential to extensive location-based services. The challenge faced with these location-based services is having to process numerous concurrent range queries over a large volume of moving objects. However, the existing range query processing algorithms are almost centralized based on one single machine, which are hard to address the challenge due to the limited memory and computing resources. To address this issue, we propose a distributed search solution for processing concurrent range queries over moving objects in this work. Firstly, a Distributed Dynamic Index (DDI) that consists of a global grid index and local dynamic M-ary tree indexes was proposed to maintain the moving objects and support the search algorithm. Next, a Distributed Range Query Algorithm (DRQA) was designed based on DDI, which introduces an incremental search strategy to monitor the range queries as objects evolve; during the process, it further designs a computation sharing paradigm for processing multiple concurrent queries by making full use of their common computation to decrease the search cost. Finally, three object datasets with different distributions were simulated on a New York road network and three baseline methods were introduced to more sufficiently evaluate the performance of our proposal. Compared with state-of-the-art method, the initial query cost of the DRQA algorithm reduces by $22.7\%$ and the incremental query cost drops by $15.2\%$, which certifies the superiority of our method over existing approaches.
\end{abstract}



\section{Introduction}

At present, location-based services are increasingly used in daily life. Continuous range query over a large volume of moving objects is essential to many location-based services. It is well-known that the car-hailing service needs to instantly find cabs within a certain range when a user submits a request and monitor the nearby cabs when the user is moving, where the user and cabs can be regarded as a query point and moving objects respectively, then the task is an essential continuous range query over moving objects. Given a set $O$ of moving objects, a query point $q_{i}$ and its query range $qr_{i}$, our goal is to search moving objects covered by $qr_{i}$ as the initial result of $q_{i}$ and continuously update the query result of $q_{i}$ as objects move. 

Continuous range queries over moving objects have been extensively studied\cite{feng2017review}, but most of them are centralized and cannot be used to handle extensive concurrent queries over a large volume of moving objects because of the limited of memory and computing resources. Hence, this work aims to address this issue in a distributed search principle designed based on a cluster of servers.

\begin{figure}[htbp]
	\centering
	\setlength{\abovecaptionskip}{0pt}
	\setlength{\belowcaptionskip}{0pt}
	\subfigure[\fontsize{7.0pt}{\baselineskip}\selectfont Grid Index ]{\label{gridIndex}\includegraphics[width=0.3\textwidth]{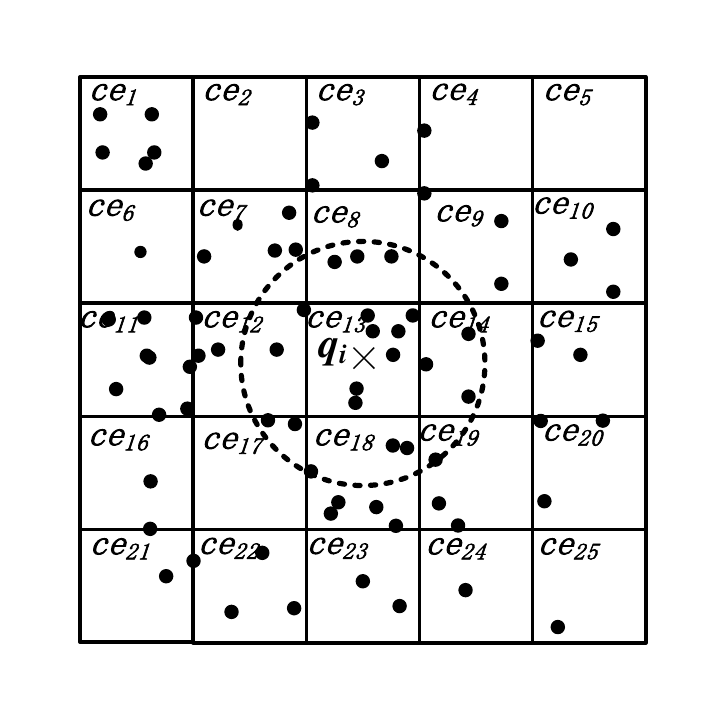}}
	\hspace{0.00in}
	\subfigure[\fontsize{7.0pt}{\baselineskip}\selectfont M-ary Tree Index ]{\label{quadtreeIndex}\includegraphics[width=0.45\textwidth]{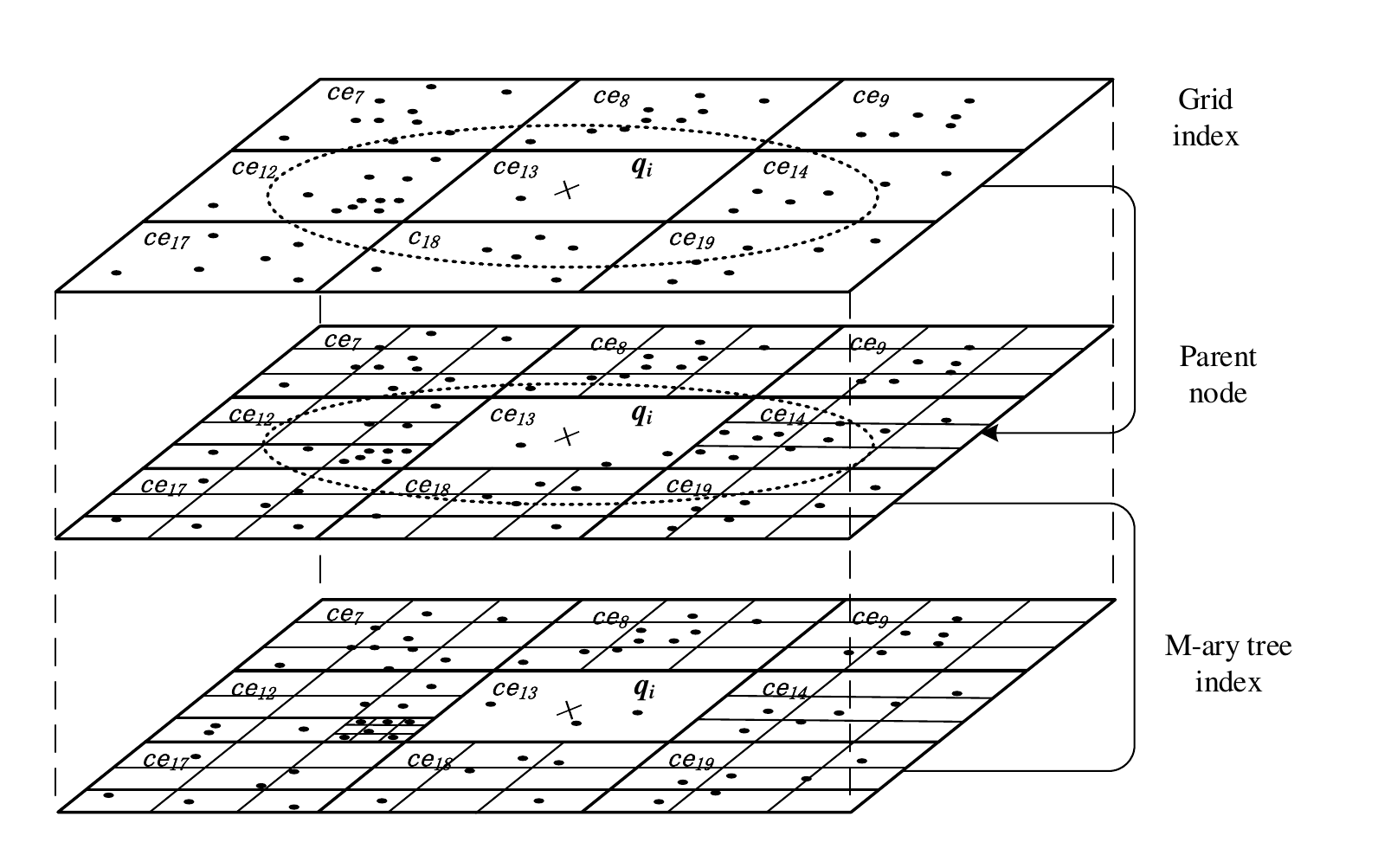}}
	\hspace{0.00in}
	\caption{DDI}		
	\label{exp:Building Time}
\end{figure}

However, the existing index structures used for processing range queries are usually R-Tree or the grid index structure, and both of them have their own flaws when deployed in a distributed environment. In particular, if we deploy a R-tree ~\cite{temporal2021,yu2019distributed} in a cluster of servers, where each server keeps a part of nodes in the R-tree. As nodes in R-tree continuously merge or split, which are expensive in a distributed environment as the merged or split nodes lie in multiple different servers. The grid index structure~\cite{kalashnikov2002efficient,tianyang2019direction,yu2005monitoring, Dong2018Direction-aware,xue2015dual,zhao2017h,li2017distributed} is easy to be deployed on a cluster of servers. However, when the algorithm based on grid index processes cells partially intersecting with the query range, it needs to scan all moving objects in each of these cells to detect which objects are covered by the query range, which will incur a great query cost when processing a large number of cells intersecting with query ranges. For example, Figure \ref{gridIndex} shows a grid index and a range query $q_{i}$, where the query range of $q_{i}$ partially overlaps with eight cells, so all moving objects in the eight cells need to be scanned and the pruning capability of the grid index is seriously weakened.


There are some existing works~\cite{cai2004processing,zhang2017distributed,hu2005generic,sun2009study,meng2016consensus} that adopt the distributed paradigm to process range queries. In the distributed paradigm, they treat each mobile device as a moving object, and the mobile devices and the data center constitute a distributed computing environment. As such, they can distribute the range query overhead to the mobile devices, but which requires the mobile devices to be powerful. However, not all mobile devices have enough computing resources, so the application of this type of approaches is limited. Moreover, frequent message transportation between mobile devices will cause considerable communication cost, which cannot be ignored and even dominate the total costs of these approaches.


In order to address above issues, we proposes a distributed dynamic index (DDI for short), which is composed of a grid index and dynamic M-ary tree indexes. The entire query area is divided into $n\times n$ cells with equal sizes, and each cell records the moving objects it contains. Each cell is independent and can be deployed to different servers. To reinforce the pruning capability of the grid index, we construct a dynamic M-ary tree for each cell, where "dynamic" means the depth of the M-ary tree can be dynamically adjusted to adapt the density of moving objects in different regions. When constructing the M-tree, each cell is regarded as a root node and $m$ child nodes are added to the root when the number of objects in the cell is greater than $\alpha$. The moving objects in the cell are stored in the leaf nodes of the M-ary tree. When a leaf node contains more than $\alpha$ objects, $m$ child nodes will be added into this leaf node and it becomes a non-leaf node. The operation will proceed like this way until no leaf node contains more than $\alpha$ objects. If a set of leaf nodes with the same parent node have less than $\beta$ objects, their parents will inherit their objects and the leaf nodes will be removed from the M-ary tree. DDI can improve the pruning power for processing range queries. An instance of DDI is shown in Figure \ref{quadtreeIndex}. As the query range of $q_{i}$ covers the entire cell $ce_{13}$, we do not need to explore its M-tree. As to each cell intersecting with the query range, we are able to explore its M-ary tree to identify the nodes that are completely covered by the query range and their objects have no need to be scanned.

Since each cell in the DDI is independent of each other, the whole index structure is easy to be deployed to distributed computing environments with Master-Worker architecture. Based on the DDI, this paper further proposes a Distributed Range Query Algorithm (DRQA for short). The algorithm decomposes each range query into sub-queries on multiple cells, which can be conducted by multiple servers in parallel to improve the query efficiency. When the range query continuously move, the DRQA incrementally computes the latest result at the current moment using the previous results of the query as much as possible. In addition, each server adopts a shared computing mechanism, which shares the computing results when processing multiple concurrent queries involving the same query region.

Main contributions can be summarized as follows:
\begin{enumerate}
    \item 
     We propose DDI that can be easily deployed in distributed computing environments and can dynamically adjust the index granularity of each cell according to the distribution density of moving objects. Moreover, DDI can help the search algorithm to reduce the traversal of unnecessary moving objects and provides greater pruning capability. 
     \item
     We propose DRQA, a Distributed Range Query Algorithm for processing extensive concurrent range queries over a large volume of moving objects. To efficiently monitoring the results of continuous range queries, an incremental search is introduced to avoid the overhaul computation in each snapshot.
     \item
     We deploy the proposed algorithm on Storm, a distributed computing platform for streaming data, and introduce three baseline methods in our experiments. Extensive experiments are conducted to verify the superiority of our proposal over the existing solutions.
\end{enumerate}

The rest of the paper is organized as follows. Section \ref{sec:related-work} briefly discusses the related work about processing range queries. Section \ref{sec:related definition} gives some related definitions. Section \ref{sec:modis} presents DDI index structure and DRQA algorithm is discussed in section \ref{sec:dsa}. The experimental results are reported in section \ref{sec:exp}. Finally, the paper is summarized in section \ref{sec:conclusion}.

\section{Related work}\label{sec:related-work}
Range query over moving objects has been widely studied as a fundamental problem in the field of location-based services. In the following, we introduce the moving object index structure, query scenarios and computational models based on the existing work.

The grid index is widely used for moving object queries~\cite{kalashnikov2002efficient,tianyang2019direction,Dong2018Direction-aware,xue2015dual,yu2005monitoring,ouyang2020progressive,shen2020grid,zhang2021sprig,li2017distributedestimate,meng2015robust}. Kalashnikov et al~\cite{kalashnikov2002efficient} proposed that the query performance based on the grid index is superior compared with other index structures. A sorting-based optimization method is also proposed to improve the cache hit rate. Dong et al~\cite{tianyang2019direction} and Tianyang Dong et al~\cite{Dong2018Direction-aware} established a moving object grid index, which can know the road network according to the orientation to determine moving objects moving to the query point. Zhongbin Xue et al~\cite{xue2015dual} proposed a memory-based high-throughput moving object range query algorithm, which can execute multiple queries at a time. Yu et al~\cite{yu2005monitoring} proposed a grid-based stretchable algorithm. The algorithm not only indexes moving objects efficiently, but also improves query performance and robustness when moving objects are densely distributed. Shen et al~\cite{shen2020grid} proposed a grid-based index structure. It expands resident areas for monitoring continuous range queries in moving or common computing environments. Zhang et al~\cite{zhang2021sprig} presented an efficient grid-based SPatial inteRpolation functIon (SPRIG) for processing the range query and the kNN query.

In order to solve the above problem, the papers~\cite{fu2017continuous,li2013brief,haidar2013finding,de2010stability,haidar2013approximate} introduced the concept of 'safety zone'. The 'safety zone' is a query movement area. It is calculated based on the distance of the nearest moving object inside or outside the query range boundary to the query boundary. In this region, the movement of the query point does not cause any change in the query result. Haidar et al~\cite{haidar2013finding,haidar2013approximate} studied a static approximate distance search algorithm, which can quickly determine the approximate range of the upper or lower bound of the 'safety zone'. Then the queries located in the specified safety zone are not reduplicate counted.

With the increasing popularity of indoor location services, the study of distance-aware queries for indoor moving objects has attracted attention from scholars~\cite{comparative2016,wang2016sme,wei2019boosting,li2020leveraging,shao2016vip,alamri2020c} in the past few years. Wang et al~\cite{wang2016sme} defined and classified moving object relationships with uncertain indoor distances. Shao et al~\cite{shao2016vip} pointed out that the features of the indoor space are underutilized by existing index structures and query algorithms. Therefore, they propose two new index structures with full consideration of indoor characteristics. They are Indoor Partitioning Tree (IP-Tree) and Vivid Indoor Partitioning Tree (VIP-Tree) respectively. Sultan et al~\cite{alamri2020c} proposed the C-tree index structure based on the grid index. C-tree can efficiently serve the indoor spatial query, the topological query, the abutting query and the density-based query.

Most of the above algorithms are centralized algorithms, which are limited by the computational capacity of the single node. Therefore, centralized algorithms are difficult to cope with large volume of concurrent queries. Silvestri et al~\cite{silvestri2014gpu} used a parallel algorithm based on the mixed use of the CPU and GPU to handle large volume of moving object range queries. In addition, several papers have studied distributed search algorithms for moving object range queries. Among them, papers~\cite{cai2004processing,zhang2017distributed,hu2005generic} adopted the idea of edge computing for range queries. Such algorithms usually require mobile terminals with strong computing power, which limits the applicability of such algorithms. In contrast, papers~\cite{yu2019distributed,fengjun2018HBase,xu2020optimization} adopted the idea of distributed computing based on server clusters. Jun Feng et al~\cite{fengjun2018HBase} proposed a static range query algorithm. The algorithm recursively traverses the road segments adjacent to the query point to update the moving objects on the road segments within the query range. However, the creation of multiple index tables imposes a huge additional load on the system when the algorithm handles large-scale concurrent queries.

To address the above problem, Jiangfeng Xu et al~\cite{xu2020optimization} proposed a multidimensional index framework New-grid. This framework uses a linearization technique based on Hilbert curves to solve this problem, and points out that the grid index has good scalability in a distributed environment. However, these algorithms still need to traverse all the moving objects in the candidate cells when dealing with candidate cells that intersect with range queries. Yu et al~\cite{yu2019distributed} proposed a hybrid distributed index structure based on a grid index and R-tree to solve the problem. The problem of the index structure is that the R-tree corresponding to each cell is constantly updated from the bottom up when the position of moving objects changes. It needs to be maintained frequently, resulting in costly index maintenance. In this paper, we believe that sharing computation results for multiple concurrent range queries involving the same region can effectively improve query efficiency.

\section{Preliminaries}\label{sec:related definition}
 Table \ref{table1} gives symbols and their meanings to be used in this paper.
 
\begin{myDef}[Moving object]
A moving object $mo_{i}$ is represented by a triple $\left \{ID_{i},t_{c},(x_{i},y_{i})\right \}$, where $ID$ is the unique identifier of $mo_{i}$, and the position of $mo_{i}$ at time $t_{c}$ is $(x_{i},y_{i})$.
\end{myDef}


\begin{myDef}[Continuous range query]
A continuous range query $q_{i}$ is represented by a triple $\left \{t_{q_{i}}^{s}, t_{q_{i}}^{e}, sq_{i}\right \}$, where $t_{q_{i}}^{s}$ is the start time of $q_i$, and $t_{q_{i}}^{e}$ is the end time of $q_i$, and $qr_i$ is the query range of $q_i$. 
\end{myDef}

In this paper, the query range $qr_{i}$ of query $q_{i}$ can be any polynomial shape. For simplicity, we assume a circle.

\begin{myDef}[Cell]
The entire search area is divided into equal-sized cells. Each cell $ce_{i}$ uses three lists $MOL_{i}$, $FCL_{i}$ and $PCL_{i}$ to record respectively the moving objects contained in cell $ce_{i}$, the query point whose query range $qr_{i}$ completely covers cell $ce_{i}$ and the query point whose query range $qr_{i}$ partially covers $ce_{i}$.
\end{myDef}

\begin{myDef}[Candidate cell]
For a range query $q_{i}$, $ce_{i}$ is a candidate cell of $q_{i}$ if the cell $ce_{i}$ completely or partially covered by $qr_{i}$. $ce_{k}\subseteq sq_{i}$ means that $ce_{k}$ is completely covered by $qr_{i}$. $ce_{k}\subset sq_{i}$ means that $ce_{k}$ is partially covered by $qr_{i}$.
\end{myDef}

\begin{table}[!htb]
\caption{Symbols Summary}\label{table1}
\begin{tabular}{c|l}
\hhline
symbol         & \multicolumn{1}{c}{meaning}                                                                                                               \\ \hline
$q_i$          & a continuous range query                                                                                                                  \\ \hline
$qr_i$         & the query range of $q_i$                                                                                                                  \\ \hline
$mo_i$          & a moving object                                                                                                                           \\ \hline
$ce_i$          & a cell                                                                                                                                    \\ \hline
$\alpha$       & \begin{tabular}[c]{@{}l@{}}the threshold for the number of moving objects \\ associated with M-ary tree splitting \end{tabular} \\ \hline
$\beta$        & \begin{tabular}[c]{@{}l@{}}the threshold for the number of moving objects \\ associated with M-ary tree merging ($\beta=\alpha/m$)\end{tabular} \\ \hline
$tn_i$          & \begin{tabular}[c]{@{}l@{}}the cell corresponds to a leaf node in the \\ M-ary tree\end{tabular}                                          \\ \hline
$ln_f$          & \begin{tabular}[c]{@{}l@{}}the cell corresponds to a leaf node in the \\ M-ary tree\end{tabular}                                          \\ \hline
$T_i$          & the M-ary tree corresponding to cell $ce_i$                                                                                                \\ \hline
$A\subset B$   & the area $A$ is partially covered by area $B$                                                                                             \\ \hline
$A\subseteq B$ & the area $A$ is completely covered by Area $B$                                                                                            \\ \hline
$mo_i\in H-Z$   & \begin{tabular}[c]{@{}l@{}}the moving object $mo_i$ belongs to set $H$, \\ but not to set $Z$\end{tabular}                                 \\ \hline
$mo_i\in tn_i$   & \begin{tabular}[c]{@{}l@{}}the moving object $mo_i$ is in the M-ary tree \\ node $tn_j$\end{tabular}                                        \\ \hline
$mo_j\in q_i$  & \begin{tabular}[c]{@{}l@{}}the moving object $mo_i$ is in the query range \\ $qr_i$\end{tabular}                                           \\ \hline
$MO_i$          & \begin{tabular}[c]{@{}l@{}}a collection of moving objects in the $qr_i$ \\ query range\end{tabular}                                       \\ \hline
$MOL_i$         & \begin{tabular}[c]{@{}l@{}}the list recording moving objects contained by \\ cell $ce_i$\end{tabular}                                      \\ \hline
$FCL_i$         & \begin{tabular}[c]{@{}l@{}}the list recording query points whose query scope \\ completely covers $ce_i$\end{tabular}                      \\ \hline
$PCL_i$         & \begin{tabular}[c]{@{}l@{}}the list recording query points whose query scope \\ partially covers $ce_i$\end{tabular}                       \\ \hline
$LN_f$         & \begin{tabular}[c]{@{}l@{}}the list recording IDs of moving objects \\ maintained by $ln_f$\end{tabular}                                   \\ \hline
$QL_i$         & \begin{tabular}[c]{@{}l@{}}the list recording query points whose query scope\\  completely covers $tn_i$\end{tabular}                      \\ \hline
$GR_i$          & \begin{tabular}[c]{@{}l@{}}a collection of candidate cells associated with \\ query $q_i$\end{tabular}                                    \\ \hline
\hhline
\end{tabular}
\end{table}

\section{Distributed index structure}\label{sec:modis}
In this section, we first discuss the structure of Distributed Dynamic Index (DDI for short) in Section \ref{grid}, and then introduce the construction of DDI in Section \ref{DDI}. In Section \ref{maintenance}, we study the maintenance of DDI as object evolve and present the distributed deployment of DDI in Section \ref{deployment}.

\subsection{Structure of DDI}\label{grid}
 We propose a distributed dynamic index (DDI for short), which consists of a grid index and dynamic M-ary trees.

 \subsubsection{Grid index}
  First, the whole search area is divided into equal-sized cells. Each cell $ce_{i}$ maintains three lists $MOL_{i}$, $FCL_{i}$, and $PCL_{i}$, which respectively record the moving objects contained in cell $ce_{i}$, the query point whose query range $qr_{i}$ completely covers cell $ce_{i}$, and the query point whose query range $qr_{i}$ partially covers $ce_{i}$. We will build a dynamic M-ary tree $T_{i}$ for the cell $ce_{i}$ when the number of moving objects in cell $ce_{i}$ exceeds the threshold $\alpha$.

 \subsubsection{Dynamic M-ary tree}
 Firstly, each leaf node $ln_f$ of the M-ary tree maintains a moving objects list $QL_{f}$, which records IDs of these moving objects covered by the leaf node. Each non-leaf node $tn_i$ in the M-ary tree maintains a query list $QL_i$, which records query IDs that the query range $qr_i$ can fully cover the node $tn_i$. Each leaf node $ln_f$ also maintains a query list $QL_i$, which records the query IDs whose query scope completely covers the leaf node or partially intersects with the leaf node. As objects move, $T_i$ can dynamically adjust the depth of the M-ary tree according to the distribution density of moving objects. Specifically, if the number of moving objects in a leaf node of $T_i$ is equal to or greater than the specified threshold, the leaf node will be recursively split until the number of moving objects in each leaf node is less than the $\alpha $. If a set of leaf nodes with the same parent node have less than $\beta$ objects, their parents will inherit their objects and the leaf nodes will be removed from the M-ary tree.

\begin{figure}[htbp]
	\centering
	\setlength{\abovecaptionskip}{0pt}
	\setlength{\belowcaptionskip}{0pt}
	\subfigure[\fontsize{7.0pt}{\baselineskip}\selectfont M-ary tree construction process ]{\label{fig2(a)}\includegraphics[width=0.4\textwidth]{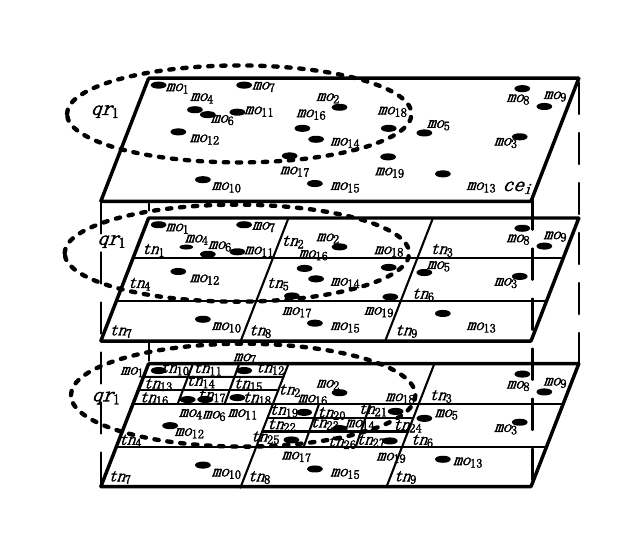}}
	\hspace{0.00in}
	\subfigure[\fontsize{7.0pt}{\baselineskip}\selectfont The structure of a M-ary tree ]{\label{fig2(b)}\includegraphics[width=8cm,height=4.5cm]{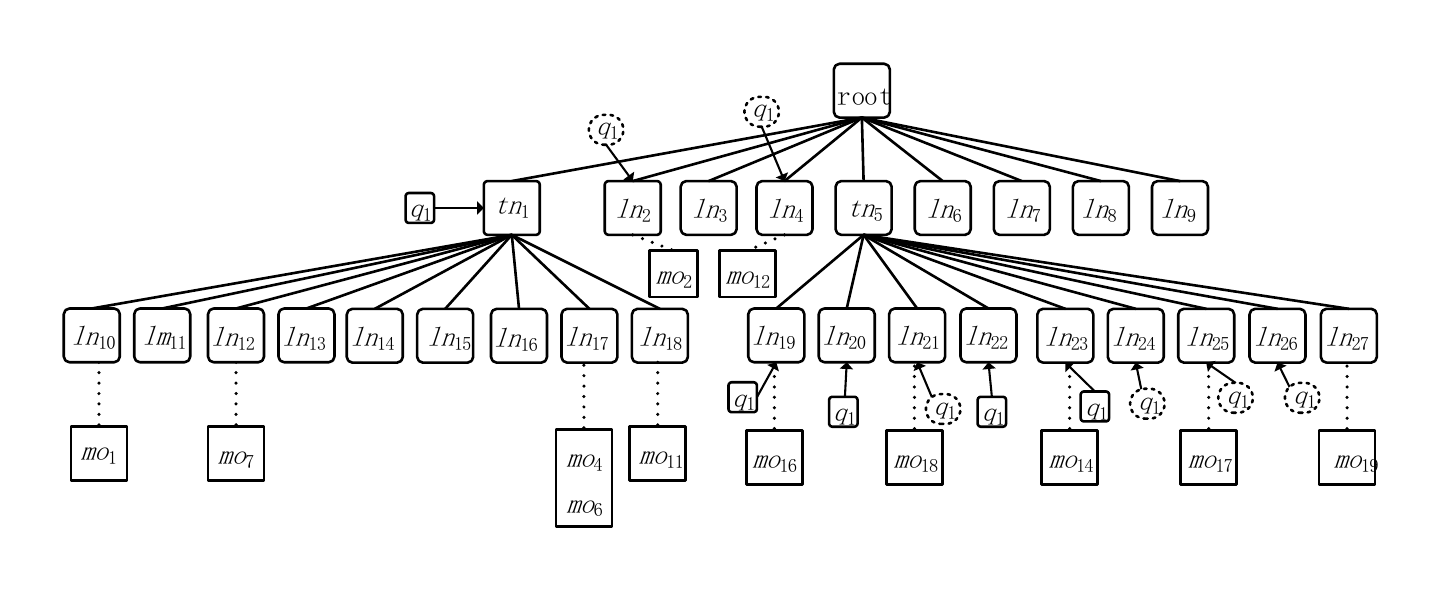}}
	\hspace{0.00in}
	\caption{Dynamic M-ary tree index}		
	\label{M-ary tree index}
\end{figure}

 Figure \ref{M-ary tree index} shows the M-ary tree index structure corresponding to cell $ce_i$. Figure \ref{fig2(a)} shows a M-ary tree construction process. Set $\alpha$=5 and $m$=9. Since the number of moving objects in cell $ce_i$ is more than $\alpha $, cell $ce_i$ is regarded as the root node and nine child nodes ($ln_1$ to $ln_9$) are added into cell $ce_i$. Since the number of moving objects in $ln_1$ and $ln_5$ is greater than $\alpha $, child nodes are respectively added into $tn_1$ and $tn_5$ until the number of moving objects in all leaf nodes is less than $\alpha $. The corresponding M-ary tree index structure is shown in Figure \ref{fig2(b)}. The query scope $qr_1$ of query $q_1$ completely covers $tn_1$, $ln_{19}$, $ln_{20}$, $ln_{22}$, $ln_{23}$. Then $q_1$ is inserted into query lists of $tn_1$, $ln_{19}$, $ln_{20}$, $ln_{22}$, $ln_{23}$. Since $qr_1$ partially intersects with leaf nodes $ln_2$, $ln_4$, $ln_{21}$, $ln_{24}$, $ln_{25}$, $ln_{26}$, the query $q_1$ is inserted into the query list of $ln_2$, $ln_4$, $ln_{21}$, $ln_{24}$, $ln_{25}$, $ln_{26}$.

The purpose of introducing the dynamic M-ary tree index structure into the grid index is to avoid traversing all moving objects in a cell when dealing with a range query that partially intersects with the cell. Specifically, if the cell $ce_i$ contains a large volume of moving objects and partially intersects with the query range $qr_i$, searching for the moving objects belonging to cell $ce_i$ and covered by $qr_i$ needs to traverse all the moving objects in the cell, which wastes a lot of computational resources. After the dynamic M-ary tree is introduced into each cell, the query algorithm can get accurate query results only by traversing partial nodes of the M-ary tree, which effectively reduces the computational cost. 

 In DDI, each cell is an independent index unit. Each cell can be deployed to any server of a distributed cluster. Each sever can maintain any number of grid indexes according to its own ability.

\subsection{Construction of DDI}\label{DDI}
Building DDI is to insert moving objects and continuous range queries into different cells as well as their corresponding M-ary trees.

\subsubsection{Object insertion}
Algorithm \ref{insert object} describes the procedure of inserting a moving object $mo_i$. Firstly, we search the current cell $ce_j$ where $mo_i$ is located. Then we add $mo_i$ into the moving object list $MOL_j$ of the cell $ce_j$. Finally, we insert $mo_i$ into the M-ary tree $T_i$ corresponds to $ce_j$. In the procedure of inserting the M-ary tree, we scan down layer by layer from the root node of $T_i$ to determine the leaf node $ln_f$ covering $mo_i (x_i, y_i)$. Then $mo_i$ is added to the moving object list $NL_f$ of $ln_f$ (lines 4-8). If the number of moving objects in the leaf node $ln_f$ reaches the $\alpha $, the leaf node $ln_f$ is split into $m$ new leaf nodes, and all moving objects in $ln_f$ are inserted into the corresponding new leaf nodes (lines 9-10).

 In algorithm \ref{insert object}, given a moving object $mo_i(x_i, y_i)$, the time to determine the cell $ce_i$ in which $mo_i$ is located based on the coordinates $(x_i, y_i)$ of the moving object $mo_i$ is $T(2n)$. We set the number of nodes in the M-ary tree corresponding to the cell $ce_i$ is $w$. It takes $T\left ( log_{4}^{w}\right )$ to determine the leaf node of the M-ary tree where $mo_i$ is located. Therefore, the time complexity of the algorithm \ref{insert object} is $T\left ( n+log_{4}^{w}\right )$.
 
\begin{algorithm}[h]
\label{insert object}
\begin{algorithmic}[1]
\REQUIRE
 a moving object $mo_i$, a M-ary tree $T_i$;
\ENSURE
the M-ary tree $T_i$;
\STATE Set a queue $Q_{p}=\text{\O}$;
\STATE Add $r_i$ into $Q_p$; 
\WHILE{$Q_{p}\neq \text{\O}$ }
\STATE $ln_f=Q_p.searchLeaf$; 
\IF{$mo_i\notin n_f$}
\STATE Continue;
\ELSIF{$mo_i\in tn_i$}
\STATE Add $mo_i$ into the moving object list $NL_f$ of $ln_f$;
\IF{$n_{f}.OL.size\geq \alpha $}
\STATE $ln_f.createChild$; 
\ENDIF
\ENDIF
\ENDWHILE
 \caption{Object insertion}\label{insert object}
 \end{algorithmic}
\end{algorithm}

\subsubsection{Continuous range query insertion}
Algorithm \ref{al:global-path} describes the process of inserting a continuous range query $q_i$. Firstly, we search for cells associated with the range query $q_i$. For the cell $ce_j$ that is completely covered by the query range $qr_i$, $q_i$ is directly added into the query list $FCL_j$ of cell $ce_j$. For the cell $ce_j$ that is partially covered by the query range $qr_i$, $q_i$ is added into the query list $PCL_j$ of cell $ce_j$. Next, we perform a top-to-bottom hierarchical traversal of the M-ary tree $T_l$ in this cell to find a M-ary tree node $tn_i$ that is completely covered by $qr_i$. $q_i$ is added into the query list $QL_i$ of node $tn_i$. And then we stop processing the child node of $tn_i$ (lines 4-9). If a node $tn_j$ in the M-ary tree $T_l$ partially intersects with the query range $qr_i$, we further traverse the child nodes of node $tn_j$ until the leaf nodes (lines 11-14). According to the insertion principle of $q_i$, only leaf nodes may partially intersect with $qr_i$ among all nodes recorded in the M-ary tree $T_l$. Therefore, all moving objects in the node area are covered by $qr_i$ if a non-leaf node of $T_l$ records $q_i$.

 In algorithm \ref{al:global-path}, the time required to determine $r$ cells intersecting with the query $q_i$ is $T(2n)$. Assuming that $r$ cells are distributed in different servers, the query $q_i$ can be inserted into the M-ary tree corresponding to different cells in parallel. At this point, the time that query $q_i$ is inserted into the DDI is equal to the maximum time that $q_i$ is inserted into a cell. Therefore, the time complexity of the algorithm \ref{al:global-path} can be expressed as $O\left ( n+Max_{1}^{r}\left ( log_{4}^{w_{i}}\right )\right )\left ( i\in \left [ 1,r\right ]\right )$, where $log_{4}^{w_{i}}$ denotes the time complexity of inserting the query $q_i$ into the M-ary tree corresponding to cell $ce_i$ and $w_i$ denotes the number of nodes in the M-ary tree.

\begin{algorithm}[h]
\begin{algorithmic}[1]
\REQUIRE
a query point $q_i$, a query range $qr_i$, a M-ary tree $T_i$;
\ENSURE
the M-ary tree $T_i$;
\STATE Set a queue $Q_{n}=\text{\O}$;
\STATE Add $r_i$ into $Q_n$; 
\WHILE{$Q_{n}\neq \text{\O}$ }
\STATE $tn_i=Q_n.nextChildNode$; 
\IF{$tn_i\subset sq_i==false$}
\STATE Continue;
\ELSIF{$tn_i\subseteq sq_i==true$}
\STATE Add $q_i$ into the query list $QL_i$ of node $tn_i$;
\STATE Continue;
\ELSIF{$tn_i.child()\neq \text{\O}$}
\STATE Add child node of $tn_i$ into $Q_n$;
\ELSE
\STATE Add $q_i$ into the query list $QL_i$ of node $tn_i$;
\ENDIF
\ENDWHILE
 \caption{Range query insertion }\label{al:global-path}
 \end{algorithmic}
\end{algorithm}

\subsection{Maintenance of DDI}\label{maintenance}
With the location of moving objects and range query points changing, the DDI needs to be maintained in real time. In the following, we discuss the maintenance of moving objects and the maintenance of range queries respectively.

\subsubsection{Maintenance of moving objects}
We suppose the moving object $mo_j$ moves from $(x_j,y_j)$ to $( x_{j}^{'},y_{j}^{'})$. For the cells and corresponding M-ary trees related with $(x_j,y_j)$ and $( x_{j}^{'},y_{j}^{'})$, the following processing should be done:

\begin{enumerate}
    \item 
     If $(x_j,y_j)$ and $( x_{j}^{'},y_{j}^{'})$ belong to the same cell $ce_o$, $mo_j$ is removed from the leaf node containing $(x_j,y_j)$. Meanwhile, $mo_j$ is added into the leaf node containing $( x_{j}^{'},y_{j}^{'})$, and the corresponding M-ary tree $T_o$ of $ce_o$ is updated.
    \item 
     If $(x_j,y_j)$ and $( x_{j}^{'},y_{j}^{'})$ belong to two different cells $ce_o$ and $ce_n$ respectively, $mo_j$ will be deleted from $MOL_o$ and leaf nodes containing $(x_j,y_j)$,  and the corresponding M-ary tree $T_o$ of $ce_o$ will be updated. Meanwhile, we add $mo_j$ into $MOL_n$ and the leaf node containing $( x_{j}^{'},y_{j}^{'})$, and update the M-ary tree $T_n$ corresponding to $ce_n$.
\end{enumerate}
 It should be noted that when updating the M-ary trees corresponding to cells in the above process, it needs to be handled according to the following conditions:

\begin{enumerate}
    \item 
    If the sum of the moving objects of the leaf node containing $(x_j,y_j)$ and its brother nodes is less than the threshold $\beta$ due to the movement of $mo_j$, all the moving objects of the leaf node in this group are saved by their parent node. Then the group of leaf nodes are deleted, and their parent node becomes a new leaf node.
    \item
    If the number of moving objects in the leaf node containing $( x_{j}^{'},y_{j}^{'})$ reaches the threshold $\alpha$ due to the movement of $mo_j$, the leaf node is recursively split until the number of moving objects in each leaf node is less than the threshold $\alpha$.
\end{enumerate}

\subsubsection{Maintenance of range query}
Suppose the range query area $qr_i$ becomes $qr_{i}^{'}$ after the query point $q_i$ is moved. At this point, the query list in the cells related to $qr_i$ and $qr_{i}^{'}$ needs to be processed according to the following:

\begin{enumerate}
    \item 
    If $qr_i$ covers $ce_i$ and $qr_{i}^{'}$ does not intersect with $ce_i$, $q_i$ is removed from the list $FCL_i$ of $ce_i$.
    \item
    If $qr_i$ covers $ce_i$ and $qr_{i}^{'}$ intersect with $ce_i$, $q_i$ is removed from $FCL_i$ and inserted into the list $PCL_i$ of $ce_i$.
    \item
     If $qr_i$ intersect with $ce_i$ and $qr_{i}^{'}$ does not intersect with $ce_i$, $q_i$ is removed from the list $PCL_i$ of $ce_i$.
     \item
     If $qr_i$ does not intersect with $ce_i$ and $qr_{i}^{'}$ covers $ce_i$, $q_i$ is inserted into the list $FCL_i$ of $ce_i$.
     \item
     If $qr_i$ intersects with $ce_i$ and $qr_{i}^{'}$ covers $ce_i$, $q_i$ is removed from $PCL_i$ and inserted into the list $FCL_i$ of $ce_i$.
     \item
     If $qr_i$ does not intersect with $ce_i$ and $qr_{i}^{'}$ intersects with $ce_i$, $q_i$ is inserted into the list $PCL_i$ of $ce_i$.
\end{enumerate}

 The maintenance process of cell $ce_i$ above only maintains the root node of $T_v$, while the maintenance of a M-ary tree $T_v$ requires top-down judging whether the query range can cover each node. And the range query is reinserted to maintain the query list $QL_i$.

 \subsection{Distributed deployment of DDI}\label{deployment}

\begin{figure*}[htbp]
\centering
\includegraphics[width=1\textwidth]{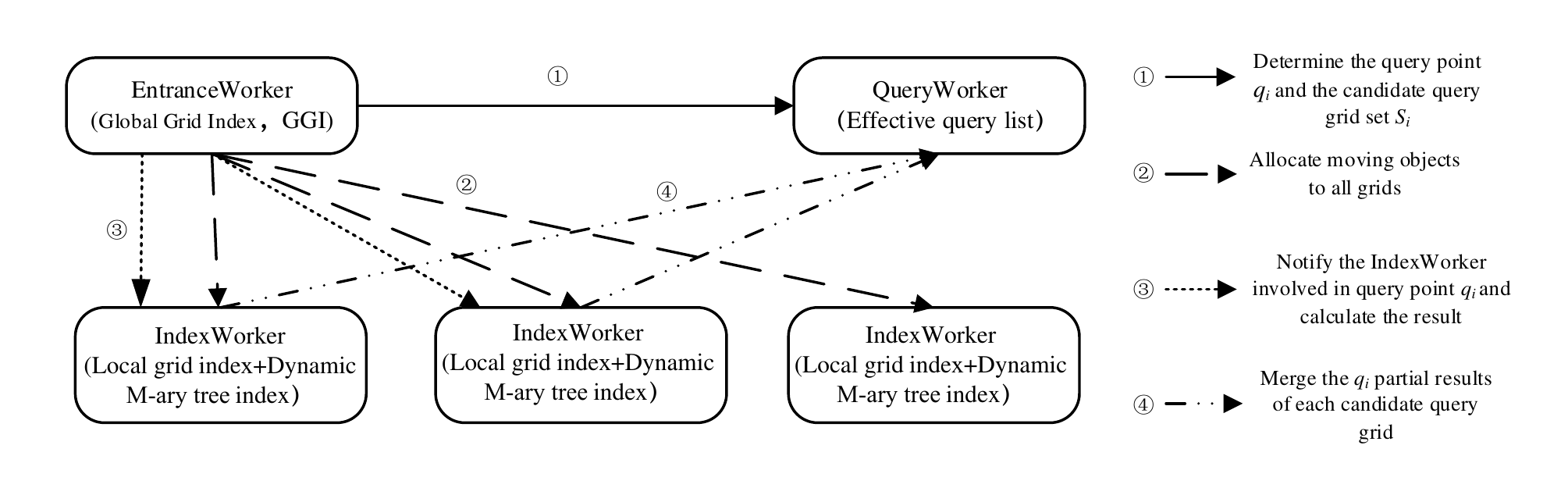}
\caption{Distributed development of DDI and DRQA}\label{fig3}
\end{figure*}

 DDI is deployed in a distributed computing environment of a Master-Worker paradigm, which consists of an EntranceWorker server, multiple IndexWorker servers and multiple QueryWorker servers. The distributed deployment framework of DDI is shown in Figure \ref{fig3}. First of all, a Global Grid Index (GGI for short) is deployed on the EntranceWorker server. GGI records the boundaries of each cell and the correspondence between all cells and IndexWorkers. Once the moving object or range query reaches the EntranceWorker server, the server determines the cells associated with the moving object or range query based on GGI and distributes them to the corresponding IndexWorker. Each IndexWorker inserts moving objects and range queries into the corresponding cells and its M-ary tree index according to the corresponding insertion algorithm, and sends the calculation result to QueryWorker, which returns the final result.

\section{Distributed Range Query Algorithm}\label{sec:dsa}
 In this section, we first describe Distributed Range Query Algorithm (DRQA for short) for moving object range queries, then further introduce the shared computing optimization strategy for concurrent queries, and finally describe the distributed incremental query algorithm in the case of moving object locations and range query locations continuous changes.

 \subsection{DRQA overview}\label{sec:philosogy-ANO}
 Distributed Range Query Algorithm (DRQA for short) for moving object range queries is proposed based on DDI. The algorithm flow is shown in Figure \ref{fig3}. For the query $q_i$, the EntranceWorker server first determines the candidate cell set $GR_i$ of query $q_i$ according to GGI, and sends the query $q_i$ and set $GR_i$ to the QueryWorker. Thereafter, the QueryWorker maintains query $q_i$. Then, the EntranceWorker server notifies multiple IndexWorkers who maintain candidate cells to search the moving objects covered by the query range $qr_i$ of $q_i$ in their respective cells in parallel. In this search process, if a candidate cell is covered by $qr_i$, all the moving objects in the cell are the query result of $q_i$. Otherwise, the M-ary tree corresponding to the candidate cell is traversed from top to bottom, so that we can quickly find the region intersecting with $qr_i$ and get the moving objects belonging to $q_i$ in the cell. Each IndexWorker sends the query results to the QueryWorker after it gets the moving objects that belonging to $q_i$ in the candidate cells under its responsibility by computing. Whenever QueryWorker receives partial query results sent by a server, it judges whether the results of all candidate cells have been received according to set $GR_i$. If all the query results have been received, the final query result of $q_i$ are returned.

 \subsection{Computation sharing mechanism}\label{sec:ANO-detail}

\begin{enumerate}
    \item 
    Transmission cost optimization. As we all know, the network transmission cost for the same scale data in a distributed environment is much greater than the CPU calculation cost. When an IndexWorker gets the query results of different queries, it needs to send the query results to QueryWorkers that maintain these queries. At this time, the more the number of QueryWorkers, the greater the network transmission cost. To solve this problem, the strategy adopted in this paper is to route multiple range queries with more overlapping query ranges to the same QueryWorker for processing. That is to say, the IndexWorker that handles these range queries needs to send the query results to the same QueryWorker. Specifically, for the two range queries $q_i$ and $q_j$, the EntranceWorker calculates the candidate cell sets of $q_i$ and $q_j$ as $GR_i$ and $GR_j$ respectively based on GGI. Then the EntranceWorker calculates the Jaccard similarity $\delta = \left | G_{i}\cap G_{j}\right |/\left |G_{i}\cup G_{j} \right |$ of sets $GR_i$ and $GR_j$. If $\delta $ is greater than or equal to the specified threshold, $q_i$ and $q_j$ are sent to the same QueryWorker for processing. Otherwise, $q_i$ and $q_j$ are sent to different QueryWorkers for processing.
    \item
    Suppose the query range $qr_i$ of query $q_i$ partially intersects with cell $ce_j$. When traversing the M-ary tree $T_j$ corresponding to cell $ce_j$ to calculate the query result of $q_i$, we find a non-leaf node of $T_j$ is completely covered by the query range $qr_i$. In order to get moving objects belonging to that non-leaf node, we still need to continue traversing down until the leaf node. Because the same cell is likely to partially intersect with multiple range queries, these range queries require multiple traversals of the same M-ary tree. This not only requires a lot of repeated calculations but also consumes a lot of time. To solve this problem, we introduce a Bipartite Graph Index (BGI for short) in each cell.
 \end{enumerate}

\begin{figure}[htbp]
	\centering
	\setlength{\abovecaptionskip}{0pt}
	\setlength{\belowcaptionskip}{0pt}
	\subfigure[\fontsize{7.0pt}{\baselineskip}\selectfont BGI structure ]{\label{fig4(a)}\includegraphics[width=0.4\textwidth]{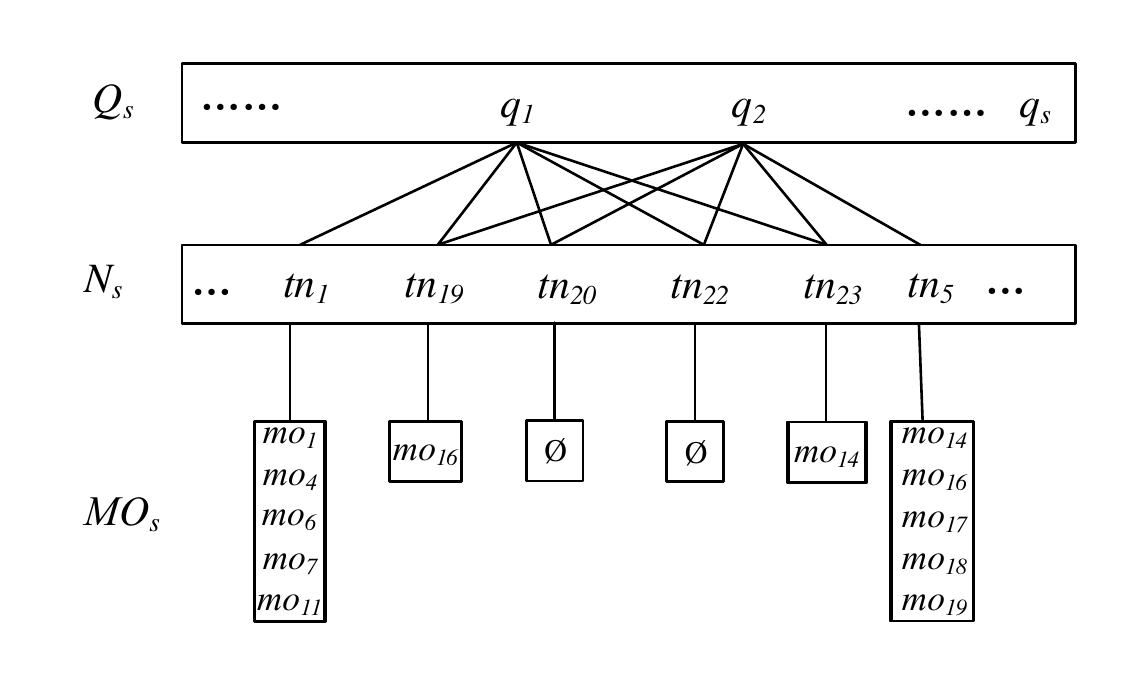}}
	\hspace{0.00in}
	\subfigure[\fontsize{7.0pt}{\baselineskip}\selectfont M-ary tree index ]{\label{fig4(b)}\includegraphics[width=0.4\textwidth]{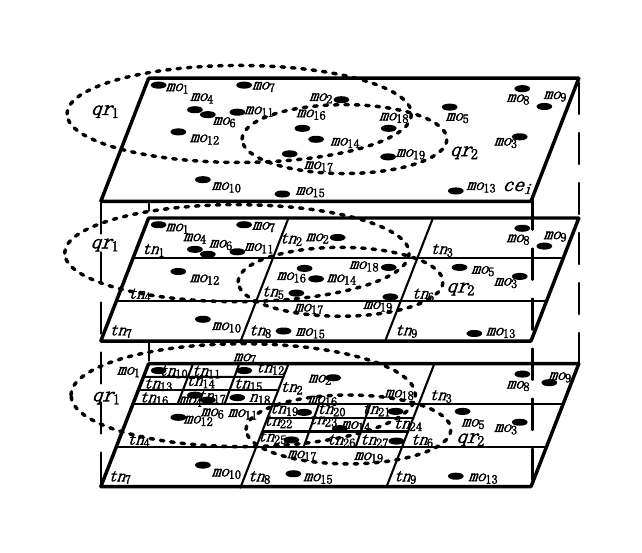}}
	\hspace{0.00in}
	\caption{Bipartite Graph Index}		
	\label{BGI}
\end{figure}

As shown in Figure \ref{fig4(a)}, BGI consists of the query set $Q_s$, the M-ary tree node set $N_s$ and the set $O_s$ of moving objects corresponding to the M-ary tree node. When querying the moving objects contained in $qr_i$, we traverse the root node of the M-ary tree $T_j$ corresponding to cell $ce_j$. If a node $tn_k$ of $T_j$ is completely covered by $qr_i$, and the query $q_j$ is recorded in the query list $QL_k$ of node $tn_k$, we find $q_j$ in the set $Q_s$ of BGI. Then we find $tn_k$ among the nodes corresponding to $q_j$ in the set $N_s$ according to $q_j$. Then we look for $tn_k$ in the set $N_s$ corresponding to $q_j$. At this point, the set $O_s$ of moving objects of $tn_k$ is determined. We don't need to continue traversing the child nodes of $tn_k$. Otherwise, we continue traversing the child nodes of $tn_k$ to obtain the set $O_s$ of moving objects of $tn_k$. Thereafter, if the query range of a query covers node $tn_k$, we can get the moving objects of $tn_k$ directly from the $N_s$ set of BGI.

To better understand the above process, we set m=9. As shown in Figure \ref{fig4(b)}, range query $qr_1$ covers nodes $tn_1$, $tn_{19}$, $tn_{20}$, $tn_{22}$, $tn_{23}$. We assume that the 9-ary tree nodes covered by the query point $qr_1$ and the moving objects in these nodes have been inserted into BGI. At this moment, if the new query range $qr_2$ and $qr_1$ both cover nodes $tn_{19}$, $tn_{20}$, $tn_{22}$, $tn_{23}$, the query result about $tn_{19}$, $tn_{20}$, $tn_{22}$, $tn_{23}$ in $qr_2$ can be obtained directly from set $N_s$. In addition, $qr_2$ covers node $tn_5$ and its corresponding moving objects are stored in the BGI after the computation is completed (as shown in Figure \ref{fig4(a)}).

 \subsection{Distributed incremental search algorithm}
  For continuous range queries for which the initial results have been obtained, this paper needs to update their query results every $\bigtriangleup t$ according to the position changes of moving objects and query points until the query fails. Therefore, this paper proposes an incremental query strategy to update the query results incrementally on the basis of making full use of the existing results. The incremental query strategy is discussed as follows.

\begin{enumerate}
    \item 
    The moving object position changes while the query point is fixed. After $\bigtriangleup t$, moving objects whose positions have changed report their previous positions and current positions to the EntranceWorker. Based on GGI, EntranceWorker determines a cell set $G$, which records the cells that moving objects need to be updated. Then EntranceWorker notifies corresponding IndexWorkers to update moving objects in set $G$. If a moving object is located in the same cell $ce_m$ before and after the move, there is no impact on queries in list $FCL_m$. Only queries in $PCL_m$ need to be updated. While updating queries in list $PCL_m$, multiple queries in list $PCL_m$ can be updated simultaneously based on BGI in cell $ce_m$. That is to say, when the position of moving objects change but the query point is fixed, the query results of query points in the query set $Q_s$ are updated according to the query list $QL_i$ of $tn_i$ if the moving object set of any node $tn_i$ in the set $N_s$ in BGI changes.
    \item
    The position of the moving object is fixed while the position of the query point changes. We assume that the query range changes from $qr_i$ to $qr_i'$ after the query $q_i$ is moved. When $q_i$ reaches the EntranceWorker, it calculates the candidate cell sets $GR_l$ and $GR_c$ of $q_i$ at the previous moment and the current moment according to $qr_i$ and $qr_i'$ respectively. If cell $ce_k$ belongs to $GR_l$ but not to $GR_c$, the IndexWorker ($IW_k$ for short) maintaining $ce_k$ notifies the QueryWorker ($QW_i$ for short) responsible for $q_i$ to remove the moving objects belonging to $ce_k$ from the query result of $q_i$ at the previous moment. If cell $ce_k$ belongs to $GR_c$ but not to $GR_l$, $IW_k$ informs $QW_i$ to add the moving object belonging to $ce_k$ to the query result of $q_i$ at the previous moment. If cell $ce_k$ belongs to both $GR_l$ and $GR_c$, $ce_k$ is processed according to the following rules.
    
    If $\left ( sq_{i}\subseteq c_{k}\right ) \&\& \left ( sq_{i}'\subseteq c_{k}\right )$ and the query range of $q_i$ always covers cell $ce_k$, the query results don't need to be updated.
    
    If $\left ( sq_{i}\subset c_{k}\right ) \&\& \left ( sq_{i}'\subseteq c_{k}\right )$, $q_i$ is removed from $PCL_k$ and inserted into $FCL_k$. $IW_k$ calculates the moving objects in region $\left ( c_{k}-\left ( c_{k}\cap sq_{i}\right )\right )$ according to the M-ary tree $T_k$. Then $IW_k$ sends these moving objects to $QW_i$, which adds these moving objects to query result of $q_i$.
    
    If $\left ( sq_{i}\subseteq c_{k}\right ) \&\& \left ( sq_{i}'\subset c_{k}\right )$, $q_i$ is removed from $FCL_k$ and inserted into $PCL_k$. $IW_k$ calculates the moving objects in region $\left ( c_{k}-\left ( c_{k}\cap sq_{i}'\right )\right )$ according to the M-ary tree $T_k$. Then $IW_k$ sends these moving objects to $QW_i$, which removes these moving objects from the query result of $q_i$.
    
    If $\left ( sq_{i}\subset c_{k}\right ) \&\& \left ( sq_{i}'\subset c_{k}\right )$, $IW_k$ calculates the moving object sets $O_d$ and $O_a$ in region $\left ( sq_{i}-\left ( sq_{i}\cap sq_{i}'\right )\right )$ and $\left ( sq_{i}'-\left ( sq_{i}\cap sq_{i}'\right )\right )$ according to the M-ary tree $T_k$. Then $IW_k$ informs $QW_i$ to remove moving objects in set $O_d$ from the query result of $q_i$, and adds moving objects in set $O_a$ to the query result of $q_i$.
 \end{enumerate}

 \subsection{Time complexity of the query algorithm}
 
 For the query $q_i$ whose query range is $qr_i$, the process contains two stages: initial query and incremental query. Then we analyze the time complexity of the initial query and the time complexity of the incremental query separately.

 In the initial query stage, it is assumed that the number of candidate cells partially intersecting with $qr_i$ is $r$. And these candidate cells are distributed in different servers. According to the distributed query idea of this paper, each candidate cell can be searched in parallel. At this time, the initial query time is equal to the maximum search time of a cell. In order to understand well, we assume that the node number of the candidate cell $ce_i$ corresponding to the M-ary tree is $w_i$. The time to determine the M-ary tree nodes that are completely covered by $qr_i$ and partially intersected with $qr_i$ is $T\left ( log_{4}^{w_{i}}\right )$. In addition, we assume that the number of M-ary tree nodes completely covered by $qr_i$ is $w_{i}^{'}$, then the number of M-ary tree nodes partially intersecting with the $qr_i$ is $w_{i}^{''}$, and all of them are leaf nodes. For each M-ary tree node that is completely covered by $qr_i$, all the moving objects in it are covered by $qr_i$. Therefore, the time to process these M-ary tree nodes is $T\left ( w_{i}^{'}\right )$. For each M-ary tree leaf node that partially intersects with the $qr_i$, the algorithm needs to traverse all the moving objects in it to determine the moving objects covered by $qr_i$. Therefore, the time to search these leaf nodes is $T\left ( w_{i}^{''}\times n_{o}\right )$. $n_{o}$ is the average of moving objects contained in each leaf node. Therefore, the processing time of a cell is $T\left ( log_{4}^{w_{i}}+w_{i}^{'}+w_{i}^{''}\times n_{o}\right )$. The time complexity of the initial stage is $O\left ( Max_{1}^{r}\left ( log_{4}^{w_{i}}+w_{i}^{'}+w_{i}^{''}\times n_{o}\right )\right ) \left ( i\in \left [ 1,r\right ]\right )$.

 The incremental query stage is divided into the following two cases.
 \begin{enumerate}
     \item 
     The position of the moving object changes while the position of the query point is fixed. The $w_{i}^{'}$ M-ary tree nodes covered by $qr_i$ and the $w_{i}^{''}$ M-ary tree nodes intersecting with $qr_i$ in each candidate cell are known. We only need to search for new moving objects located in the M-ary tree nodes intersecting with $qr_i$. And these moving objects are covered by $qr_i$. The time required for the above process is $T\left ( w_{i}^{''}\times n_{o}^{e}\right )$. $n_{o}^{e}$ is the average of newly arrived moving objects in each M-ary tree node intersecting with $qr_i$. Therefore, the time complexity of the incremental query in this case is $O\left ( Max_{1}^{r} \left ( w_{i}^{''}\times n_{o}^{e}\right )\right ) \left ( i\in \left [ 1,r\right ]\right )$.
     \item
     The position of the query point changes while the position of the moving object is fixed. Due to the change of the query point location, the candidate cells need to be redetermined. The time spent on this process is $T(2n)$. At this time, we assuming that the number of candidate cells intersecting with the $qr_i$ part is $r$, the time complexity in this case is $O\left ( Max_{1}^{r}\left ( log_{4}^{w_{i}}+w_{i}^{'}+w_{i}^{''}\times n_{o}\right )\right ) \left ( i\in \left [ 1,r\right ]\right )$. Although the time complexity in this case is the same as that of the initial query, the values of $w_{i}^{''}$ and $w_{i}^{''}$ are significantly smaller than their values in the initial query. Therefore, the query efficiency is still significantly improved.
 \end{enumerate}

\section{Experiments}\label{sec:exp}

In this section, the performance of DDI and DRQA are tested. In Section \ref{dataset}, this paper describes the data set and experimental environment required by the experiment. In Section \ref{evddi} and \ref{evdsa}, this paper evaluates the performance of DDI and DRQA respectively. Finally, comparing DRQA with three baseline algorithms in Section \ref{compari}.

\subsection{Experimental setup}\label{dataset}



In this paper, the DDI is evaluated by data sets with different moving objects distribution. Three moving object data sets are simulated based on New York road network: uniform distribution (UD) data set, gaussian distribution (GD) data set and Zipf data set. To simulate these data sets, the first is to regard the two-dimensional space of cover every road network as a $1\times 1$ square area, and divided into $100\times 100$ cells with the same size. Then, the second is to calculate the number of moving objects in each cell according to the probability density distribution function. Finally, let the moving objects in each cell move at speed $V_{0}$ on the corresponding local road network. We take GD as an example to illustrate the process of data set generation. Because the coordinate $\left ( x_{i},y_{i}\right )$ of the object $mo_{i}$ in GD data set satisfies Gaussian Distribution in both $X$ dimension and $Y$ dimension, the probability $f$ that the object $mo_{i}$ located at $\left ( x_{i},y_{i}\right )$ is the probability density distribution function of all moving objects in the whole query area. Assuming that the total number of moving objects is $N_0$, then the number of moving objects in the cell of row $i$ and column $j$ should be $f\ast N_{0}$. Then, corresponding number of moving objects are generated in the cell to move continuously on the local road network covered by the cell.

This paper introduces three distributed algorithms as baseline methods to compare and evaluate the performance of DRQA. The first baseline algorithm naive search (NS) does not use any indexes, and each server contains all moving objects, and each query is randomly assigned to all servers. The second baseline algorithm grid index (GI) is to build a distributed grid index for moving objects. A given query is allocated to the corresponding servers based on the distributed grid index. Multiple servers compute the results of a given query in parallel. The third baseline algorithm DHI~\cite{yu2019distributed} is a hybrid index composed of global grid and R-tree structure proposed by Yu et al. Based on the distributed grid index, R-tree is established with the initial position of moving objects as leaf nodes, and the index granularity is adjusted by setting the size of leaf nodes.

The experimental server cluster consists of 20 Elastic Cloud Server(ECS) rented by Ali Cloud. Each server has 4 cores and 16G memory. In the experiment, parameters involved in DDI and DRQA are shown in Table \ref{table2}.

\begin{table}[htpb]
\caption{Summary of symbols Used in Evaluation}\label{table2}
\begin{tabular}{c|l}
\hhline
symbol   & \multicolumn{1}{c}{meaning}                                                                                                               \\ \hline
$\alpha$ & \begin{tabular}[c]{@{}l@{}}the threshold for the number of moving objects \\ associated with M-ary tree splitting\end{tabular} \\ \hline
$\beta $ & \begin{tabular}[c]{@{}l@{}}the threshold for the number of moving objects \\ associated with M-ary tree merging ($\beta=\alpha/m$)\end{tabular} \\ \hline
$m$      & each split of a tree node requires $m$ children to be split                                                                               \\ \hline
$v_o$    & the movement rate of the moving object                                                                                                    \\ \hline
$Q_o$    & \begin{tabular}[c]{@{}l@{}}a queue used to cache moving objects waiting to \\ be processed\end{tabular}                                   \\ \hline
$v_q$    & \begin{tabular}[c]{@{}l@{}}the rate at which a query point or moving object \\ enters the cache queue\end{tabular}                        \\ \hline
$Q_q$    & \begin{tabular}[c]{@{}l@{}}a queue used to cache query points waiting to be \\ processed\end{tabular}                                     \\ \hline
$N_o$    & moving object quantity(PCS)                                                                                                               \\ \hline
$N_q$    & query quantity (PCS)                                                                                                                      \\ \hline
$t$      & time (seconds); For example: t second                                                                                                     \\ \hline
$r$      & query radius                                                                                                                              \\ \hline
\hhline
\end{tabular}
\end{table}

\subsection{Performance of DDI}\label{evddi}
In this section, we will evaluate the building cost, maintenance cost, and throughput of DDI.

\subsubsection{Construction cost evaluation}
\begin{figure}[htbp]
	\centering
	\setlength{\abovecaptionskip}{0pt}
	\setlength{\belowcaptionskip}{0pt}
	\subfigure[\fontsize{7.0pt}{\baselineskip}\selectfont Impact of $m$ on DDI construction time ]{\label{m}\includegraphics[width=0.4\textwidth]{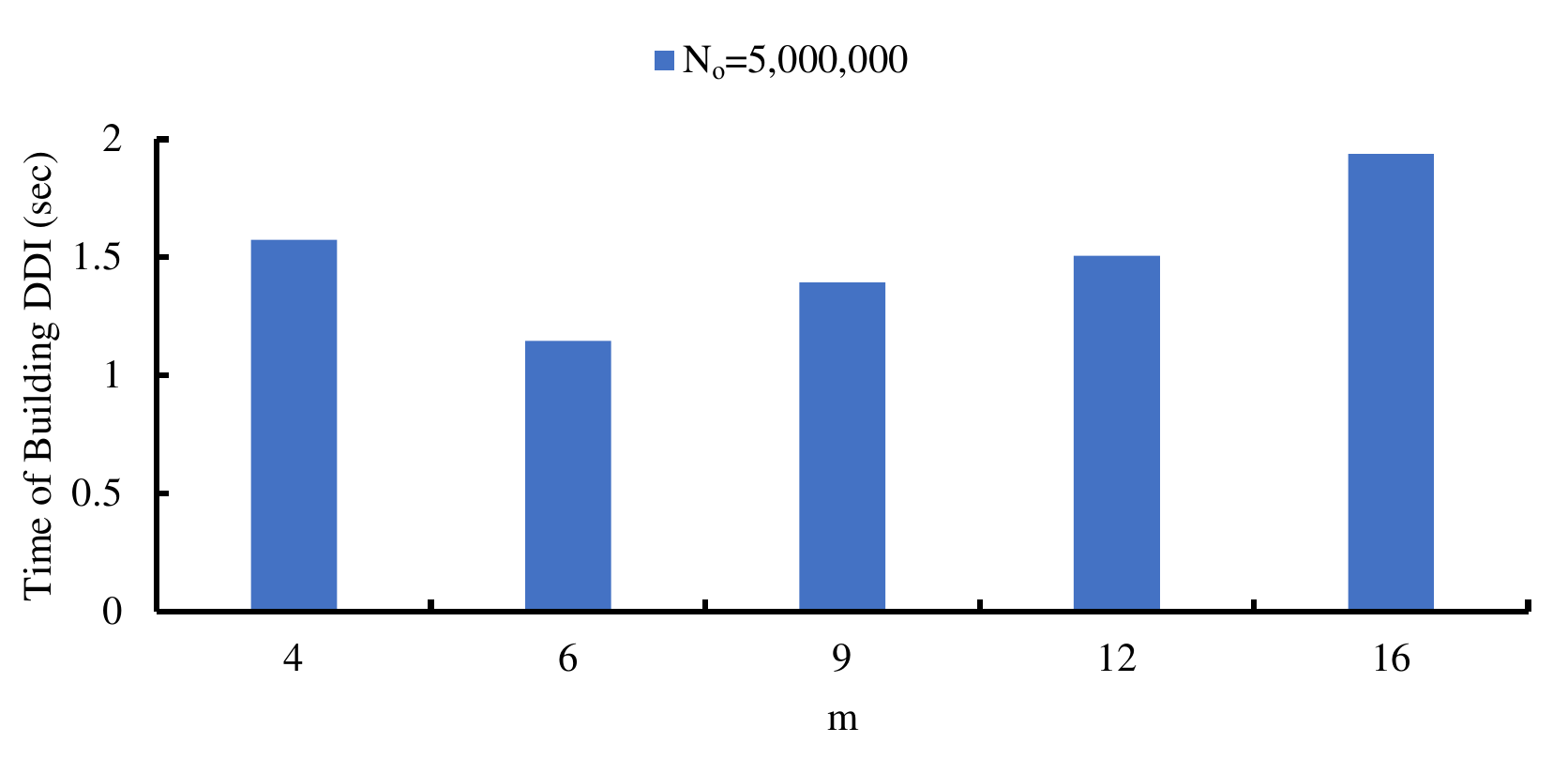}}
	\hspace{0.00in}
	\subfigure[\fontsize{7.0pt}{\baselineskip}\selectfont Impact of number of moving objects on DDI construction time ]{\label{number}\includegraphics[width=0.4\textwidth]{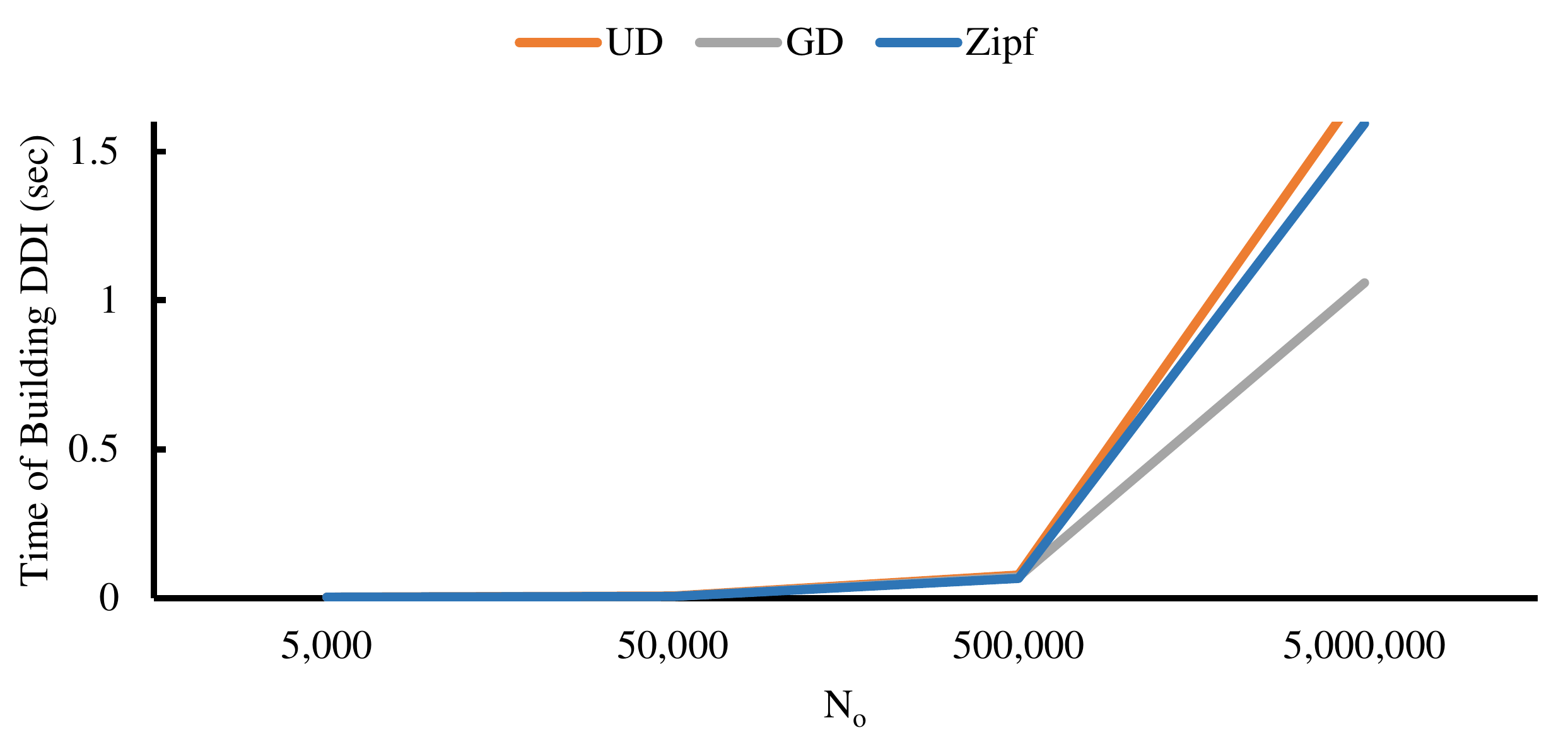}}
	\hspace{0.00in}
	\subfigure[\fontsize{7.0pt}{\baselineskip}\selectfont Impact of $\alpha $ on DDI construction time ]{\label{threshold}\includegraphics[width=0.4\textwidth]{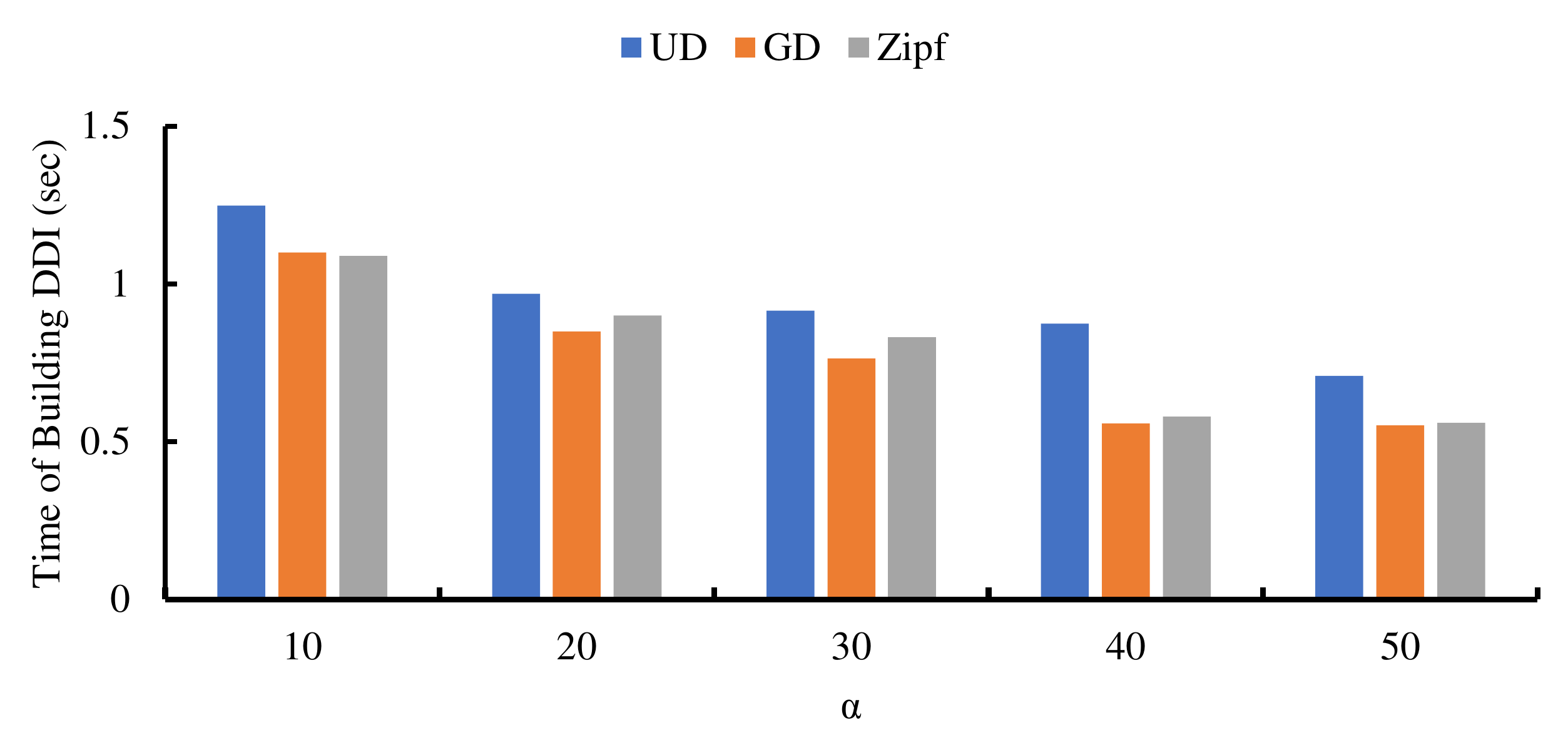}}
	\hspace{0.00in}
	\caption{Impacts of different parameters on index building time}
\end{figure}

This group of experiments tests the effect of different bifurcation values $m$ on the DDI construction time, and the experimental results are shown in Figure \ref{m}. From the experimental results, it can be seen that the DDI construction time decreases and then increases with an increase of $m$. When $m=6$, the DDI construction time is the least. The reasons are as follows. When $m<6$, although fewer child nodes need to be built for each split, more tree nodes need to be split. When $m>6$, although fewer tree nodes need to be split, more child nodes need to be created for each split as $m$ increases, so the DDI construction time increases.


In this group of experiments, moving objects were put into $Q_o$ at one time, and the time of constructing DDI for these moving objects was counted. As shown in Figure \ref{number}. The experimental results show that the DDI construction time and the number of moving objects basically increase in the same proportion. In addition, the time required to process moving objects in the GD data set is slightly lower. This is because the number of cells that need to build a M-ary tree is reduced, so the construction time decreases.

 The parameter $\alpha $ directly affects the DDI construction time. Experimental results are shown in Figure \ref{threshold}, regardless of which distribution the moving objects conform to, the construction cost of DDI decreases significantly as $\alpha $ increases. The reason is that $\alpha $ is directly related to the number of levels of the M-ary tree. The larger parameter $\alpha $, the less likely the M-ary tree will split child nodes. Thus the number of layers of the M-ary tree is smaller, which makes the construction cost of DDI smaller.

\subsubsection{Maintenance cost evaluation}
\begin{figure}[htbp]
	\centering
	\setlength{\abovecaptionskip}{0pt}
	\setlength{\belowcaptionskip}{0pt}
	\subfigure[\fontsize{7.0pt}{\baselineskip}\selectfont Influence of $m$ on the Index Maintenance Time ]{\label{fig6a}\includegraphics[width=0.4\textwidth]{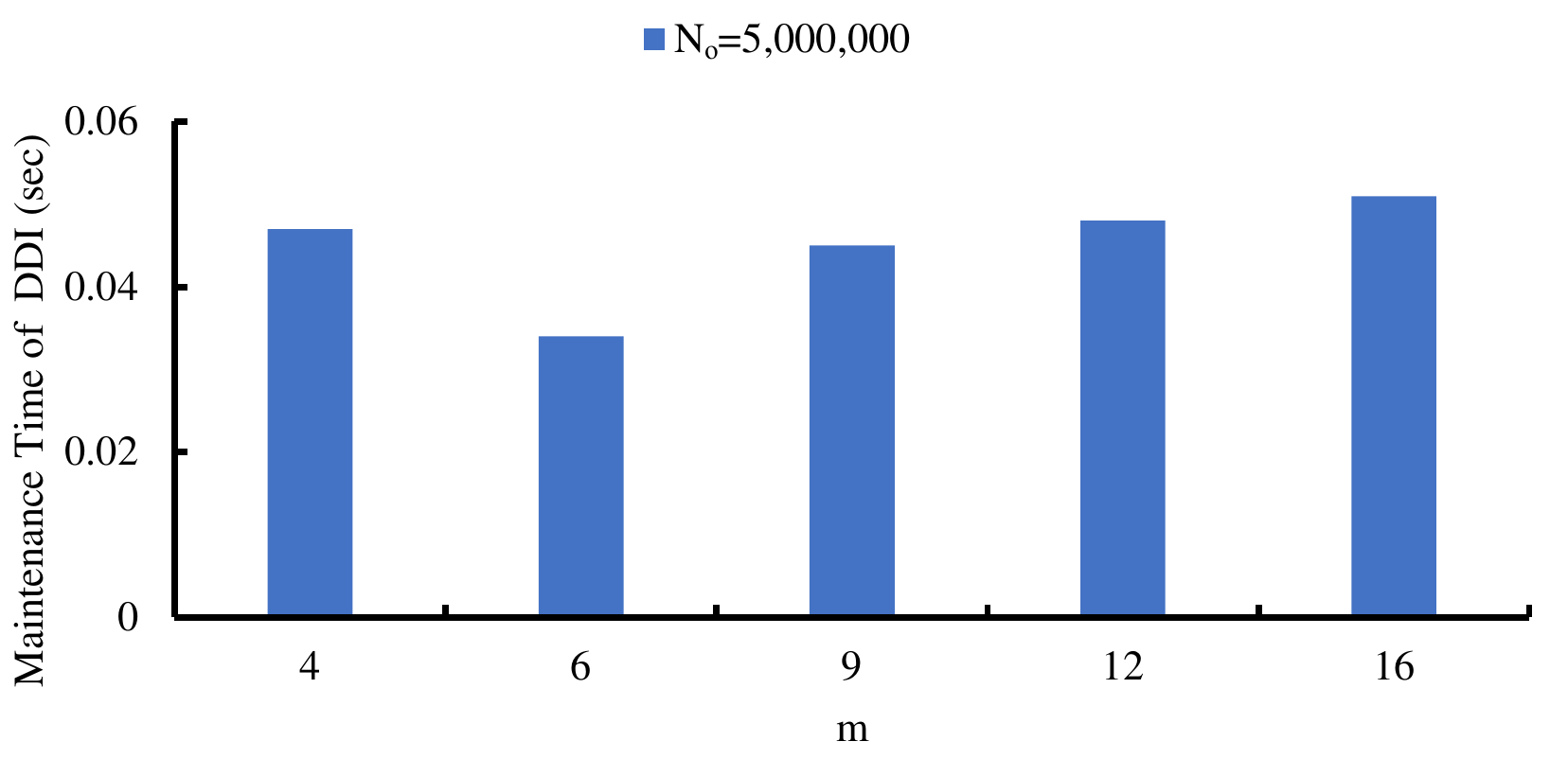}}
	\hspace{0.00in}
	\subfigure[\fontsize{7.0pt}{\baselineskip}\selectfont Influence of $v_o$ on the Index Maintenance Time ]{\label{fig6b}\includegraphics[width=0.4\textwidth]{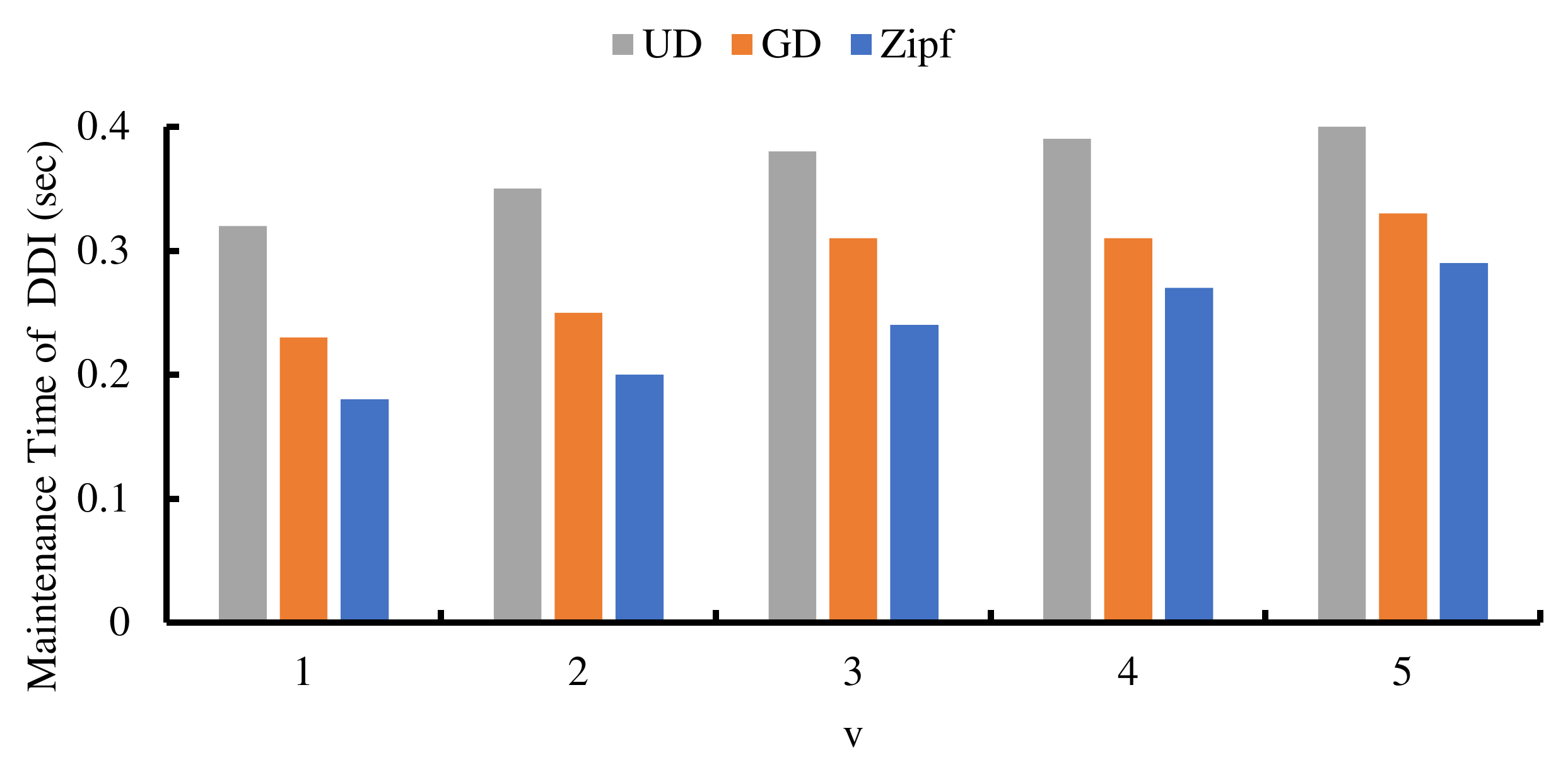}}
	\hspace{0.00in}
	\caption{Impacts of different parameters on index maintenance time}
\end{figure}

This group of experiments tests the effect of different bifurcation values $m$ on the DDI maintenance time, and the experimental results are shown in Figure \ref{fig6a}. From the experimental results, it can be seen that the DDI maintenance time decreases and then increases with the increase of $m$. When $m=6$, the maintenance time is the least. The reasons are as follows. When $m<6$, although fewer child nodes need to be traversed for each down layer, more layers need to be traversed. When $m>6$, although the number of levels to be traversed decreases, the number of child nodes to be traversed increases at each lower level as $m$ increases, so the DDI maintenance time increases.

This group of experiments first established DDI for $50M$ moving objects. Then $5M$ moving objects were randomly selected. They move five times at different rates$(v_o)$ to test the average maintenance time of DDI. The experimental results are shown in Figure \ref{fig6b}. The results show that the maintenance cost increases significantly with the increase of moving object rate. This is because when the moving rate is low, the last position and the current position of the moving object may be located in the same leaf node of the M-ary tree, and the index structure of the M-ary tree does not need to be adjusted at this time. However, when moving objects move too fast, M-ary tree nodes need to adjust the number of moving objects frequently, which increases the maintenance cost.

\subsubsection{Throughput of DDI}
\begin{figure}[htbp]
\centering
\includegraphics[width=0.4\textwidth]{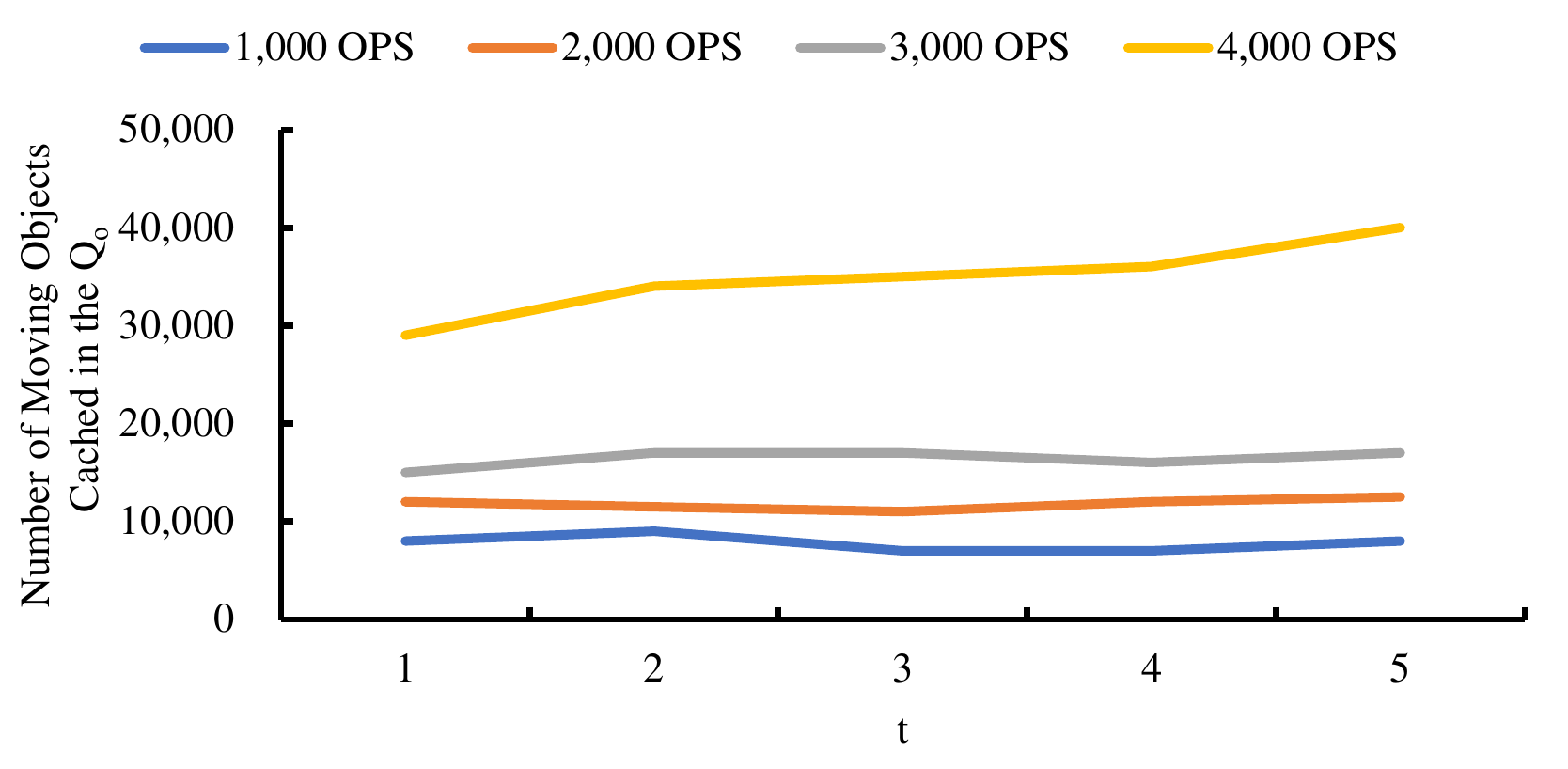}
\caption{Throughput of DDI}\label{fig7}
\end{figure}

The throughput test results for DDI are shown in Figure \ref{fig7}. In this group of experiments, we set a queue $Q_o$ to cache the moving objects to be processed. The capacity of the queue $Q_o$ is set to $50,000$. Let the moving objects feed the queue $Q_o$ at different rates. After a period of time, detect the number of moving objects $(N_o)$ in the queue $Q_o$. When the rate is less than 4 million objects per second(OPS), the number of moving objects in the queue $Q_o$ remains stable. When the rate reaches 4 million OPS, the number of moving objects in the queue $Q_o$ increases rapidly. That is to say, the throughput of DDI is about 4 million OPS. This shows that DDI can perform range query work for large volume of moving objects.

\subsection{Query Performance}\label{evdsa}
\begin{figure}[htbp]
	\centering
	\setlength{\abovecaptionskip}{0pt}
	\setlength{\belowcaptionskip}{0pt}
	\subfigure[\fontsize{7.0pt}{\baselineskip}\selectfont Impact of $m$ on query time ]{\label{fig8a}\includegraphics[width=0.4\textwidth]{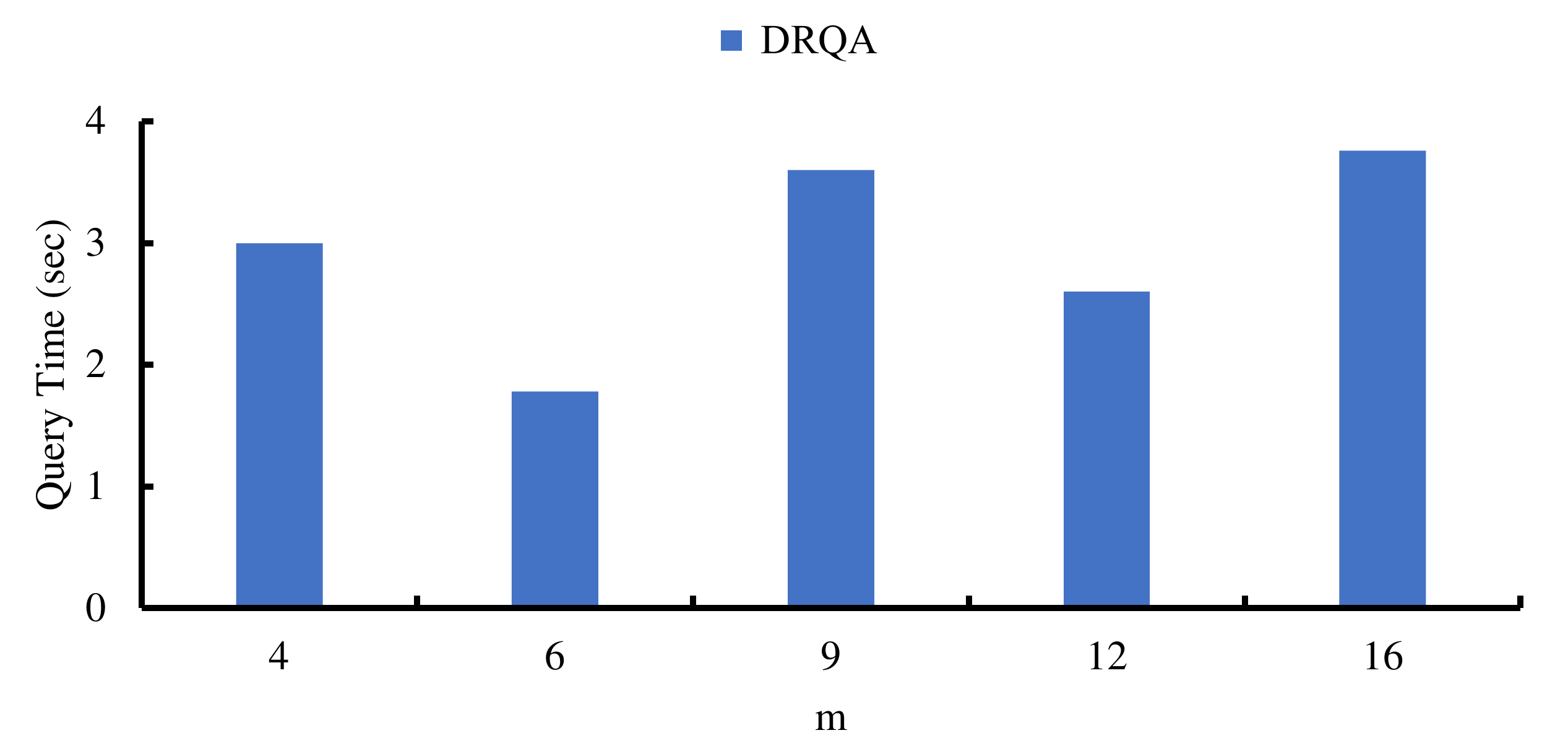}}
	\hspace{0.00in}
	\subfigure[\fontsize{7.0pt}{\baselineskip}\selectfont Impact of $\alpha $ on query time ]{\label{fig8b}\includegraphics[width=0.4\textwidth]{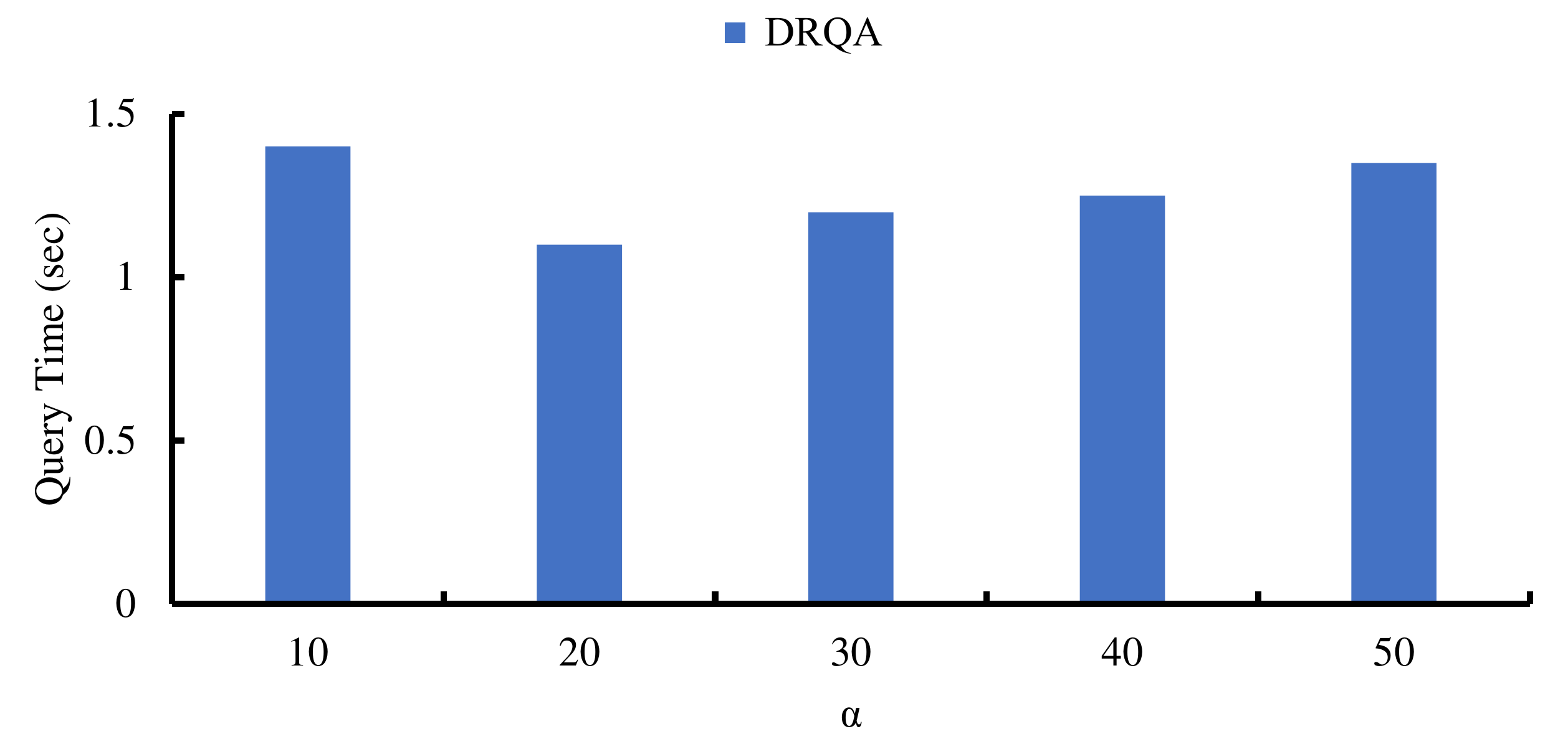}}
	\hspace{0.00in}
	\subfigure[\fontsize{7.0pt}{\baselineskip}\selectfont Impact of query range $r$ on query time ]{\label{fig8c}\includegraphics[width=0.4\textwidth]{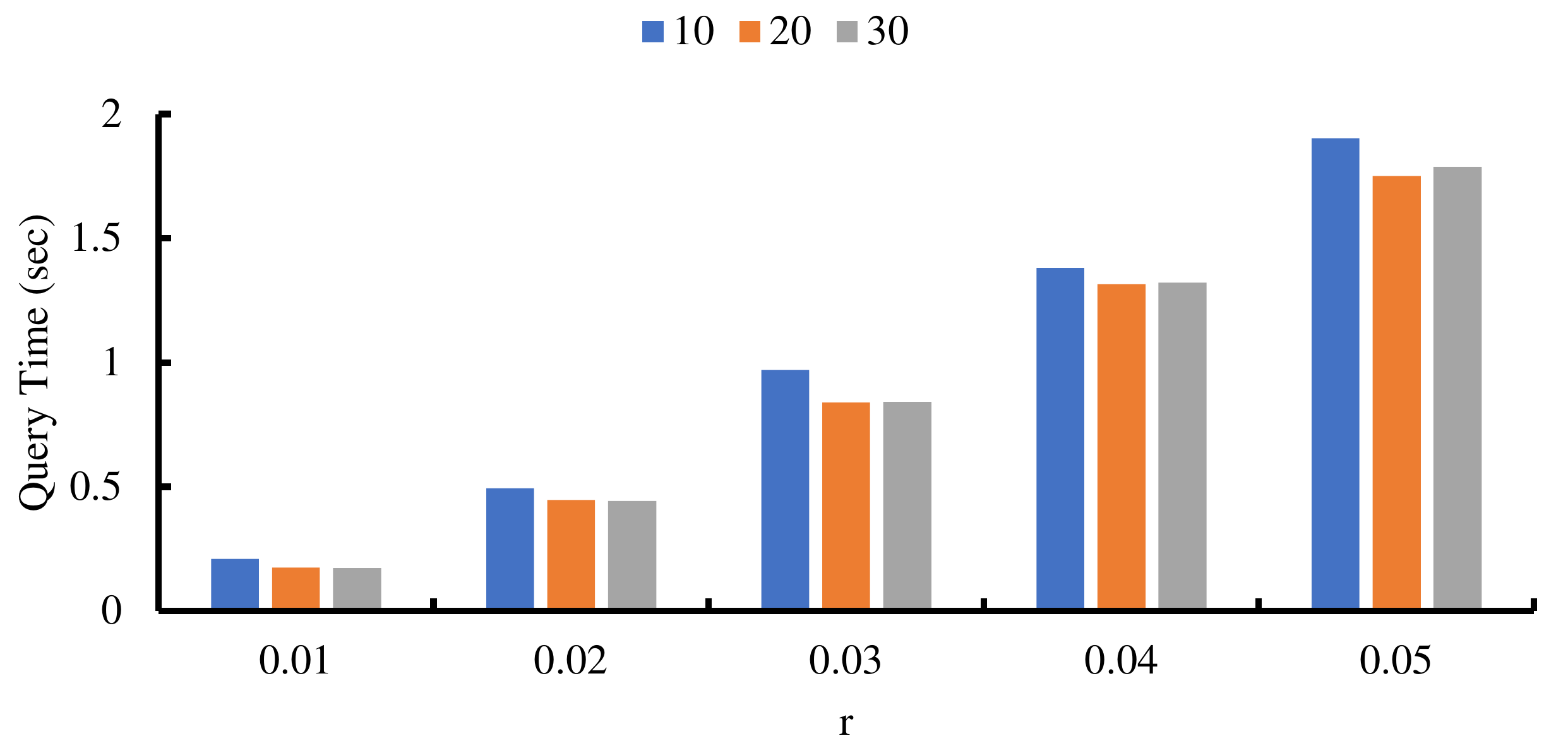}}
	\hspace{0.00in}
	\caption{Influence of different parameters on query time}
\end{figure}

This group of experiments tests the effect of different bifurcation values $m$ on the query time, and the experimental results are shown in Figure \ref{fig8a}. From the experimental results, it can be seen that the query time decreases and then increases with the increase of $m$. The best result is achieved when $m=6$. The reason is as follows. When the bifurcation value $m$ is too small, although there are fewer child nodes to be traversed in each down layer, the number of layers to be traversed is more. Conversely, when the bifurcation value $m$ is too large, although the number of levels to be traversed decreases, the number of child nodes to be traversed increases at each next level, resulting in the increase of query time. 

In this section, we test the query time of the same group of queries under different $\alpha $, and the experimental results are shown in Figure \ref{fig8b}. The results show that with the increase of the $\alpha $, the query time decreases and then increases, and the optimal effect is achieved when the $\alpha $ is set to 20. When the $\alpha $ is too small, the M-ary tree index granularity generally decreases, and the traversal cost is relatively large, resulting in the increase of in the overall query cost. On the contrary, when the threshold $\alpha $ is too large, the M-ary tree index granularity generally increases and the pruning effectiveness decreases, resulting in the increase of query time.

In this group of experiments, the DDI index with a size of $100\times 100$ was first established (the size of each cell was defined as $0.01\times 0.01$) and the relationship between DRQA performance and query range was evaluated. The results are shown in Figure \ref{fig8c} that the size of the query range has a certain influence on the performance of DRQA. The larger the query radius $r$ of a query point, the greater the number of candidate query cells involved. Even though DRQA can use multiple servers to conduct parallel queries on the moving objects of candidate cells, with the increase of the number of candidate cells, the number of servers to be processed by the system also increases correspondingly. As all the servers need to communicate with QueryWorker, the query time of DRQA increases.

\subsection{Comparison with baselines}\label{compari}
\begin{figure}[htbp]
	\centering
	\setlength{\abovecaptionskip}{0pt}
	\setlength{\belowcaptionskip}{0pt}
	\subfigure[\fontsize{7.0pt}{\baselineskip}\selectfont Impact of the number of queries on query time]{\label{fig9(a)}\includegraphics[width=0.4\textwidth]{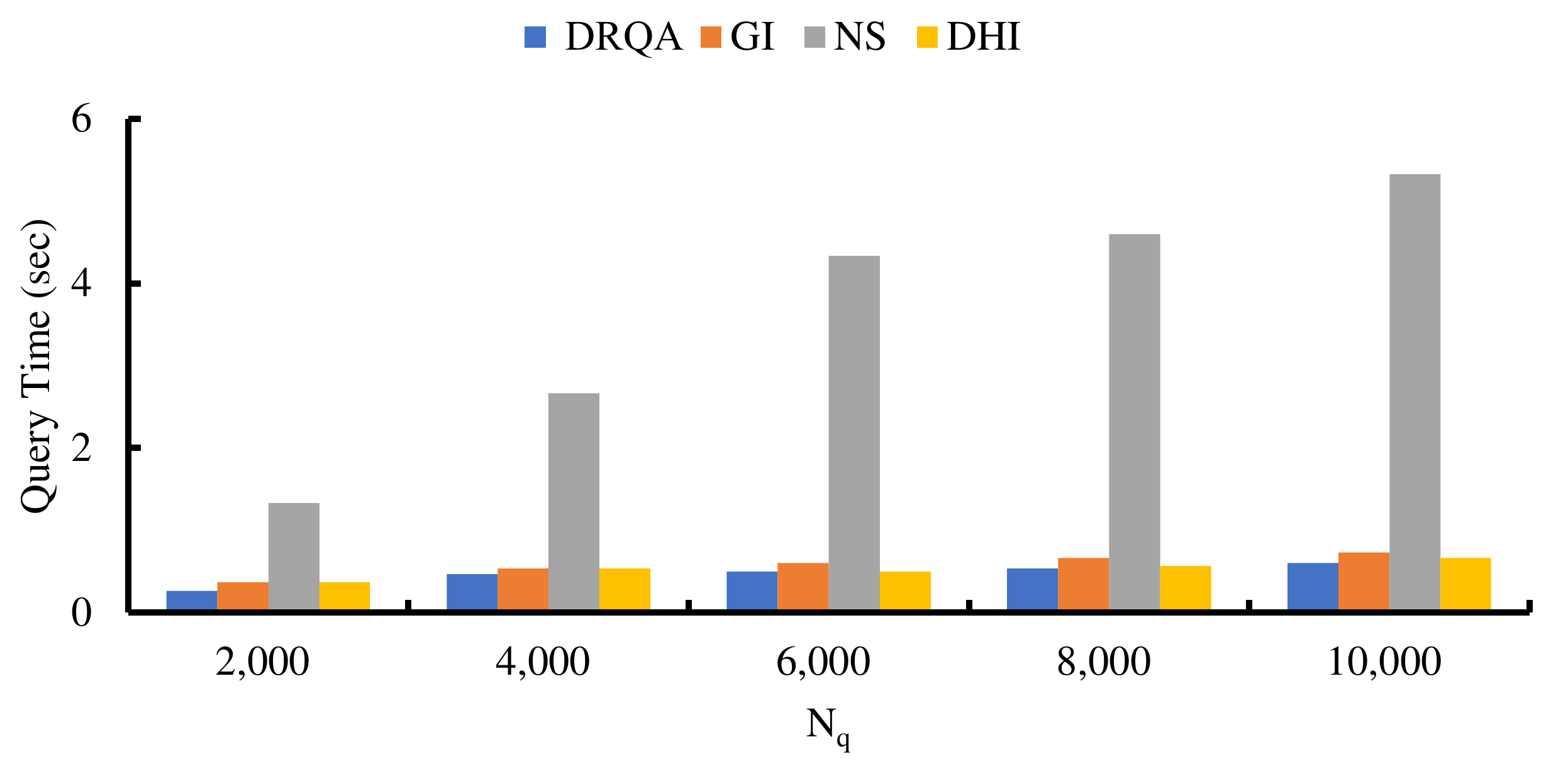}}
	\hspace{0.00in}
	\subfigure[\fontsize{7.0pt}{\baselineskip}\selectfont Impact of the number of moving objects on query time]{\label{fig9(b)}\includegraphics[width=0.4\textwidth]{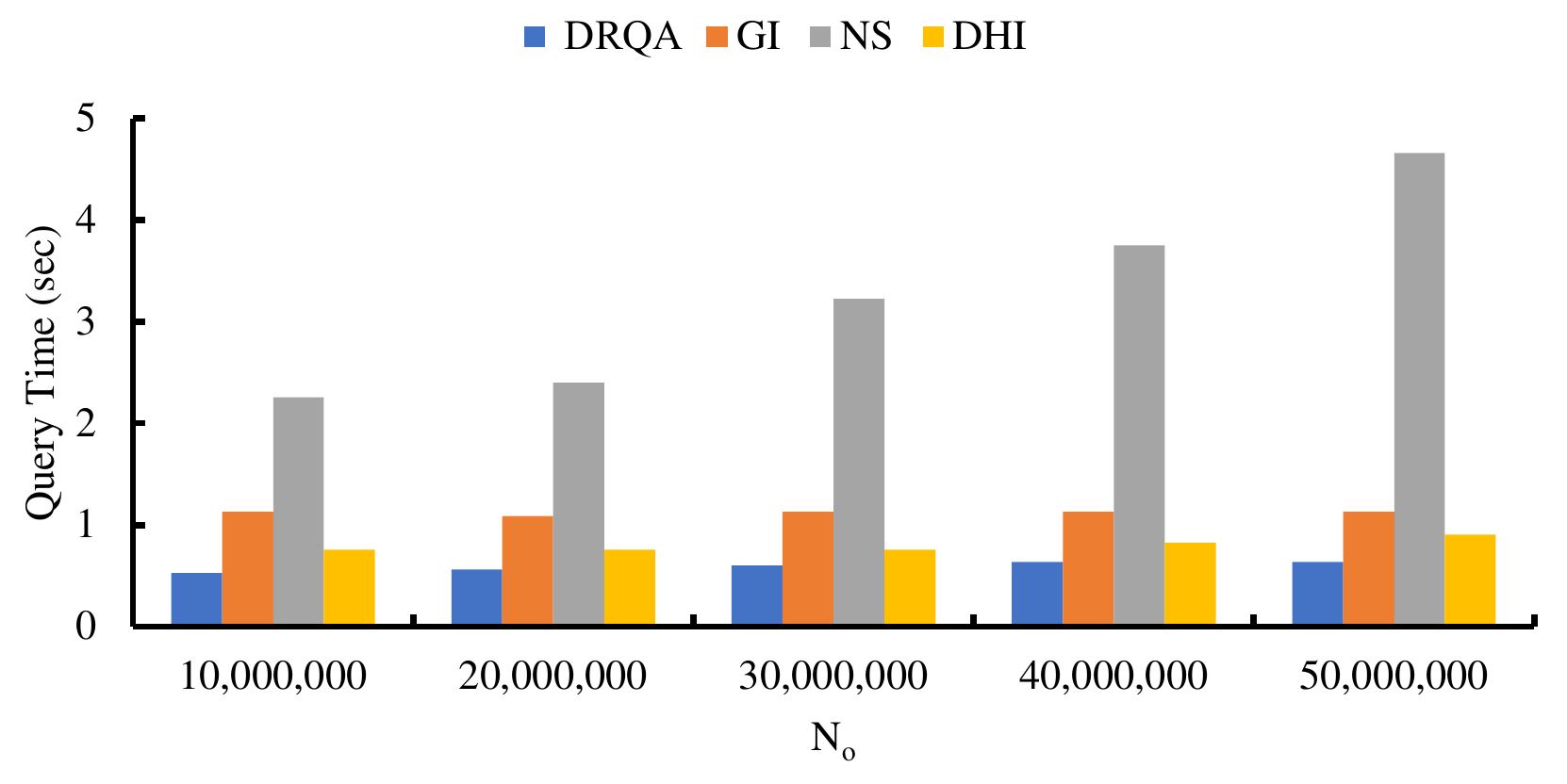}}
	\hspace{0.00in}
	\caption{Performance comparison of initial query time of multiple algorithms}		
	\label{parameters}
\end{figure}

In this group of experiments, different numbers of range queries were set to test the query time of the four algorithms. The experimental results are shown in Figure \ref{fig9(a)}. The results show that the performance of DRQA is always better than NS, GI and DHI. This is because all servers use brute force search mode when NS processes each query, which causes its query time to increase linearly with the number of continuous range queries. GI needs to detect more moving objects than DRQA due to lack of M-ary tree index support when processing the same set of contiguous range queries. During query processing, DHI fails to consider the queries with a high query area overlap rate and may allocate these queries to different servers, resulting in extra computing overhead. DRQA not only considers the optimization of transmission cost, but also adopts BGI structure to eliminate the need to traverse the bottom of the M-ary tree for multiple range queries.

In this group of experiments, different numbers of moving objects are taken as variables to test the change of query time of the four algorithms. In this group of experiments, 10,000 continuous range queries were input to $Q_q$ at one time. The processing time of all queries by the four algorithms is shown in Figure \ref{fig9(b)}. As the number of queries increases, NS query time increases linearly, while GI, DRQA, and DHI performance are almost unaffected. This is because NS needs to scan all the moving objects to process each query, whereas GI, DRQA, and DHI can effectively prune the search space so that even if the number of moving objects increases dramatically, only a limited number of moving objects need to be processed. The reason why the performance of DRQA and DHI is significantly better than GI is that DRQA and DHI introduce M-ary tree index and R-tree index respectively, which further improve the index efficiency and can effectively process large volume of moving object range queries.

\begin{figure}[htbp]
\centering
\includegraphics[width=0.4\textwidth]{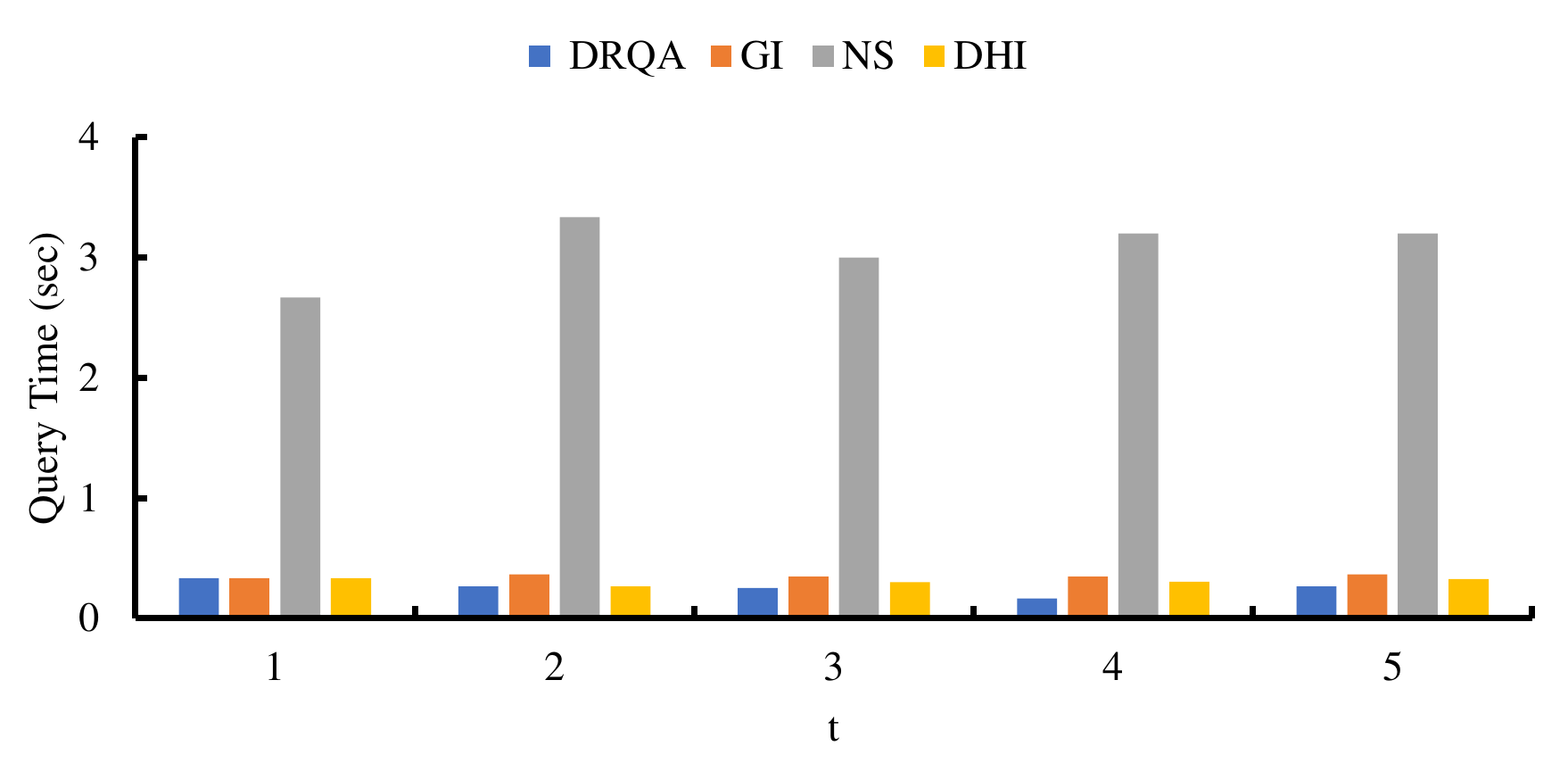}
\caption{Comparison of Incremental Query Times}\label{fig10}
\end{figure}

Firstly, 2000 range queries are fed into $Q_q$ at once and these range queries move randomly. Then we test the incremental query times of four algorithms. The experimental results are shown in Figure \ref{fig10}. It is found that the performance of DRQA is better than that of other three algorithms, especially at the last four time points. NS and GI treat each range query as a new query, and the incremental query time does not change. DHI does not consider the problem of shared computation among multiple concurrent range queries, so as to carry out the redundant computation. DRQA only computes the incremental results of each query, effectively reducing the incremental query time.

\subsubsection{Throughput comparison of different algorithms}
\begin{figure}[!htb]
	\centering
	\setlength{\abovecaptionskip}{0pt}
	\setlength{\belowcaptionskip}{0pt}
	\subfigure[\fontsize{7.0pt}{\baselineskip}\selectfont DRQA throughput]{\label{fig11(a)}\includegraphics[width=0.4\textwidth]{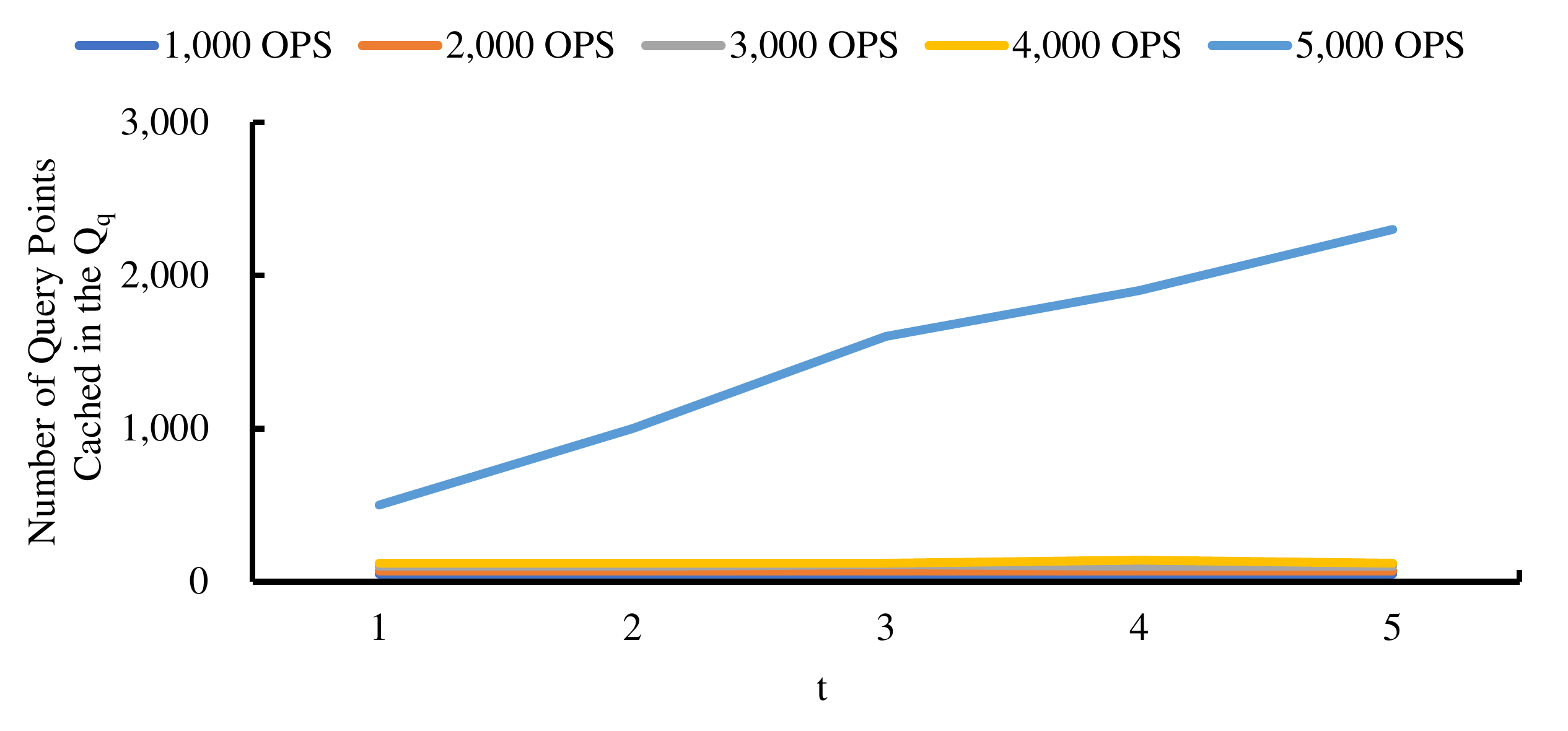}}
	\hspace{0.00in}
	\subfigure[\fontsize{7.0pt}{\baselineskip}\selectfont GI throughput]{\label{fig11(b)}\includegraphics[width=0.4\textwidth]{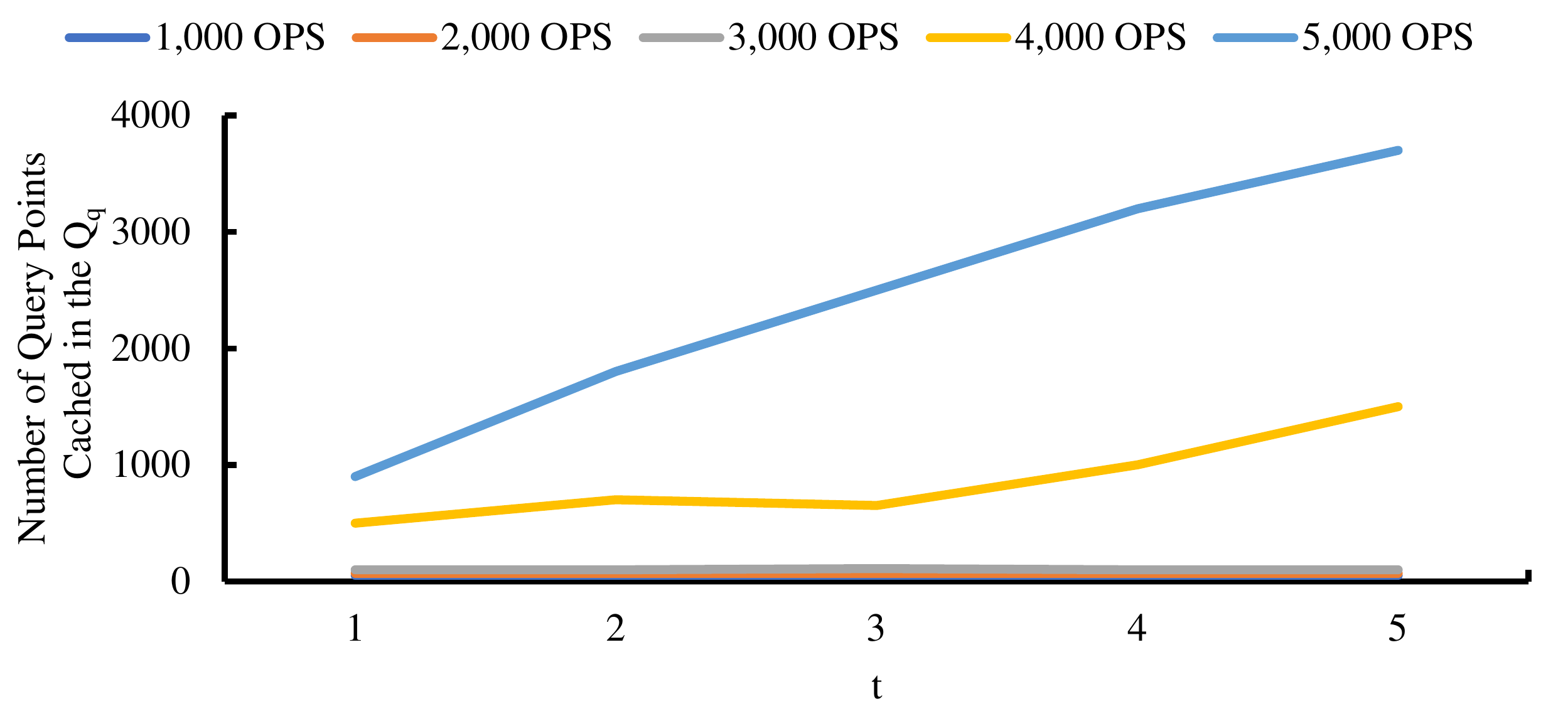}}
	\hspace{0.00in}
	\subfigure[\fontsize{7.0pt}{\baselineskip}\selectfont NS throughput]{\label{fig11(c)}\includegraphics[width=0.4\textwidth]{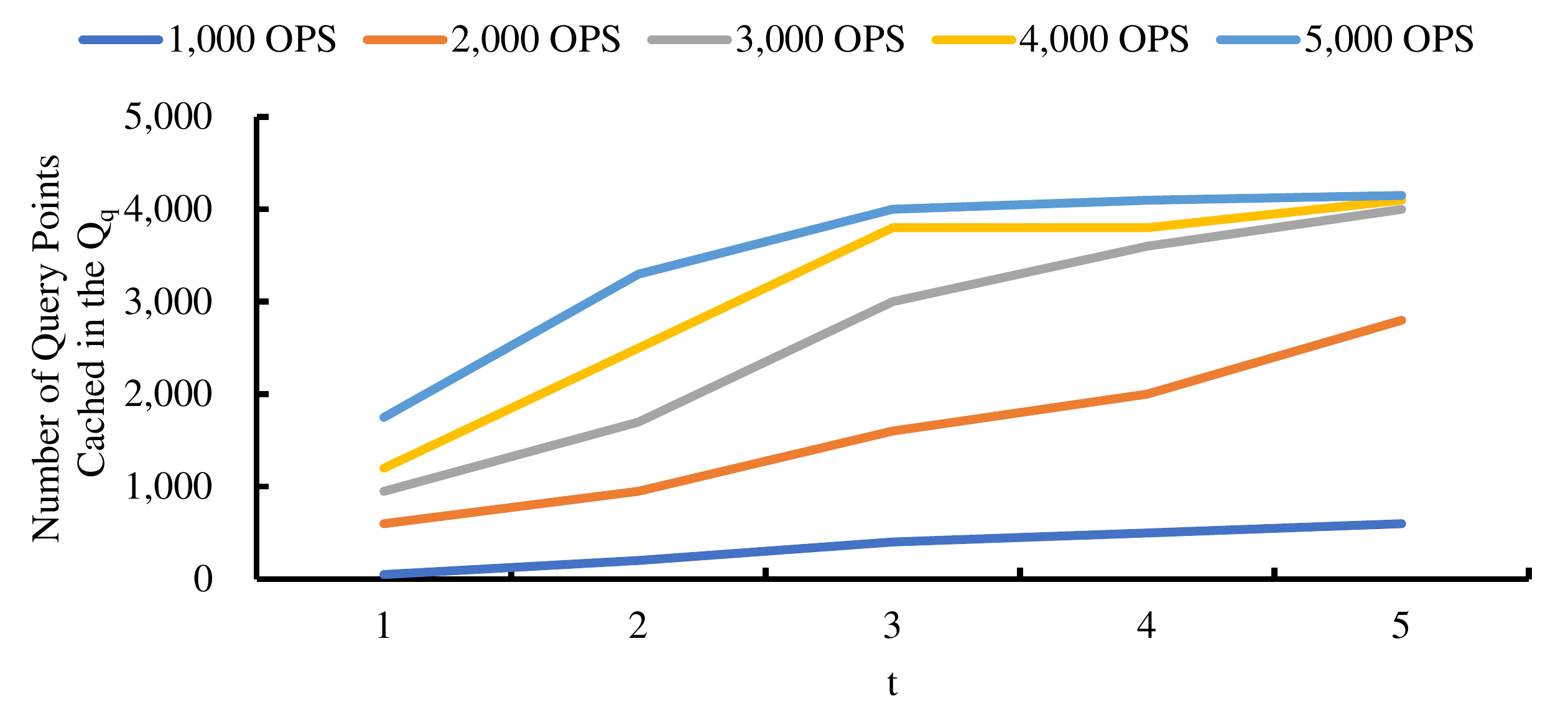}}
	\hspace{0.00in}
	\caption{Throughput of DRQA, GI and NS}		
	\label{fig11}
\end{figure}

Finally, we evaluate the throughput of the three query methods by calculating the initial results of continuous range queries. In this group of experiments, query points enter the system at different rates. Then update the number of query points cached in $Q_q$ and set the $Q_q$ cache queue capacity to 10000. Figure \ref{fig11} shows the experimental results of DRQA, GI and NS respectively. Observed from Figure \ref{fig11(a)}, the number of query points in $Q_q$ is always small when the rate $(v_q)$ is less than 4000 per second. However, the number of query points in $Q_q$ begins to increase significantly when the rate $(v_q)$ reaches 5000 per second. Therefore, the throughput of DRQA is about 4000 per second. It can be seen that the throughput of GI and NS is about 3000 per second and 1000 per second respectively from Figure \ref{fig11(b)} and Figure \ref{fig11(c)}, both smaller than DRQA. As a result, DRQA can effectively process large volume of continuous range queries over moving objects in the distributed environment.

 \balance
\section{Conclusion}\label{sec:conclusion}
 This paper proposes a distributed index structure to support continuous range queries over moving objects. The index structure has efficient pruning ability, low maintenance cost, and is easy to be deployed to a cluster of servers. Based on the index structure, we further propose a distributed incremental continuous query algorithm, which introduces the incremental search strategy and a computation sharing mechanism to enhance the efficiency of processing continuous range queries. In the future, we plan to study the distributed processing of range queries over moving objects with road network constraints.

\bibliographystyle{unsrt}
\bibliography{reference}
\end{sloppypar}
\end{document}